\documentclass[pdflatex,sn-mathphys]{sn-jnl}

\usepackage{xcolor}
\usepackage{comment}

\usepackage{amsmath}
\usepackage{amssymb}
\usepackage{bm}

\jyear{2021}%

\begin{document}

\title[Article Title]{The nature of dynamic local order in CH$_3$NH$_3$PbI$_3$ and CH$_3$NH$_3$PbBr$_3$}

\author*[1,2]{\fnm{Nicholas} {J} \sur{Weadock}}\email{nicholas.weadock@colorado.edu}
\equalcont{These authors contributed equally to this work.}
\author[3]{\fnm{Tyler} {C} 
\sur{Sterling}}
\equalcont{These authors contributed equally to this work.}

\author[4,5]{\fnm{Julian} {A} \sur{Vigil}}

\author[5]{\fnm{Aryeh} \sur{Gold-Parker}}

\author[5]{\fnm{Ian} {C} \sur{Smith}}

\author[6]{\fnm{Ballal} \sur{Ahammed}}

\author[7]{\fnm{Matthew} {J} \sur{Krogstad}}

\author[8]{\fnm{Feng} \sur{Ye}}

\author[9,10]{\fnm{David} \sur{Voneshen}}

\author[{11}]{\fnm{Peter} {M} \sur{Gehring}}

\author[12]{\fnm{Hans-Georg} \sur{Steinr{\"u}ck}}

\author[6]{\fnm{Elif} \sur{Ertekin}}

\author[5,13]{\fnm{Hemamala} {I} \sur{Karunadasa}}

\author*[3]{\fnm{Dmitry} \sur{Reznik}}\email{dmitry.reznik@colorado.edu}

\author*[1,2,14]{\fnm{Michael} {F} \sur{Toney}}\email{michael.toney@colorado.edu}

\affil[1]{\orgdiv{Materials Science and Engineering}, \orgname{University of Colorado, Boulder}, \orgaddress{\city{Boulder}, \state{CO}, \postcode{80309}, \country{USA}}}

\affil[2]{\orgdiv{Department of Chemical and Biological Engineering}, \orgname{University of Colorado, Boulder}, \orgaddress{\city{Boulder}, \state{CO}, \postcode{80309}, \country{USA}}}

\affil[3]{\orgdiv{Department of Physics}, \orgname{University of Colorado, Boulder}, \orgaddress{\city{Boulder}, \state{CO}, \postcode{80309}, \country{USA}}}

\affil[4]{\orgdiv{Department of Chemical Engineering}, \orgname{Stanford University}, \orgaddress{\city{Stanford}, \state{CA}, \postcode{94305}, \country{USA}}}

\affil[5]{\orgdiv{Department of Chemistry}, \orgname{Stanford University}, \orgaddress{\city{Stanford}, \state{CA}, \postcode{94305}, \country{USA}}}

\affil[6]{\orgdiv{Department of Mechanical Science \& Engineering and Materials Resesarch Laboratory}, \orgname{University of Illinois at Urbana-Champaign}, \orgaddress{\city{Urbana}, \state{IL}, \postcode{61801}, \country{USA}}}

\affil[7]{\orgdiv{Advanced Photon Source}, \orgname{Argonne National Lab}, \orgaddress{\city{Lemont}, \state{IL}, \postcode{60439}, \country{USA}}}

\affil[8]{\orgdiv{Neutron Scattering Division}, \orgname{Oak Ridge National Laboratory}, \orgaddress{\city{Oak Ridge}, \state{TN}, \postcode{37830}, \country{USA}}}

\affil[9]{\orgdiv{ISIS Facility}, \orgname{Rutherford Appleton Laboratory}, \orgaddress{\city{Chilton}, \state{Didcot}, \postcode{Oxon OX11 0QX}, \country{United Kingdom}}}

\affil[10]{\orgdiv{Department of Physics}, \orgname{Royal Holloway University of London}, \orgaddress{\postcode{Egham TW20 0EX}, \country{United Kingdom}}}

\affil[11]{\orgdiv{NIST Center for Neutron Research}, \orgname{National Institute of Standards and Technology}, \orgaddress{\city{Gaithersburg}, \state{MD}, \postcode{20899}, \country{USA}}}

\affil[12]{\orgdiv{Department Chemie}, \orgname{Universt{\"a}t Paderborn}, \orgaddress{\city{Paderborn}, \country{Germany}}}

\affil[13]{\orgdiv{Stanford Institute for Materials and Energy Sciences}, \orgname{SLAC National Accelerator Laboratory}, \orgaddress{\city{Menlo Park}, \state{CA}, \postcode{94025}, \country{USA}}}

\affil[14]{\orgdiv{Renewable and Sustainable Energy Institute (RASEI)}, \orgname{University of Colorado, Boulder}, \orgaddress{\city{Boulder}, \state{CO}, \postcode{80303}, \country{USA}}}

\abstract{Hybrid lead halide perovskites (LHPs) are a class of semiconductor with novel properties that are distinctively governed by structural fluctuations. Diffraction experiments sensitive to average, long-range order reveal a cubic structure in the device-relevant, high-temperature phase. Local probes find additional short-range order with lower symmetry that may govern the structure-function relationships of LHPs. However, the dimensionality, participating atoms, and dynamics of this short-range order are unresolved, impeding our understanding of technologically relevant properties including long carrier lifetimes and facile halide migration. Here, we determine the true structure of two prototypical hybrid LHPs, CH$_3$NH$_3$PbI$_3$ and CH$_3$NH$_3$PbBr$_3$, using a combination of single-crystal X-ray and neutron diffuse scattering, neutron inelastic spectroscopy, and molecular dynamics simulations. The remarkable collective dynamics we found are not suggested by previous studies and consist of a network of local two-dimensional, circular pancake-like regions of dynamically tilting lead halide octahedra (lower symmetry) that induce longer range intermolecular correlations within the CH$_3$NH$_3^+$ sublattice. The dynamic local structure can introduce transient ferroelectric or antiferroelectric domains that increase charge carrier lifetimes, and strongly affect the halide migration, a poorly understood degradation mechanism. Our approach of co-analyzing single-crystal X-ray and neutron diffuse scattering data with MD simulations will provide unparalleled insights into the structure of hybrid materials and materials with engineered disorder.}




\maketitle
The crystal structure and associated symmetry of a material are key determinants of mechanical, electronic, optical, and thermal properties. One has to look no further than seminal condensed matter physics textbooks for derivations of these properties made possible by consideration of the long-range or average order determined by the crystal lattice and translational symmetry \cite{ashcroftSolidStatePhysics2000, kittelIntroductionSolidState2019}. However, in many materials including disordered rocksalts and intercalation compounds used for battery cathodes, relaxor ferroelectrics, thermoelectrics, and oxide and halide perovskites, the  important properties are not well described by the long-range structure. Instead it is short-range order that dominates aspects of the structure-function relationship \cite{clementCationdisorderedRocksaltTransition2020, krogstadReciprocalSpaceImaging2020a, simonovHiddenDiversityVacancy2020a, xuThreedimensionalMappingDiffuse2004, krogstadRelationLocalOrder2018, rothSimpleModelVacancy2020a, lanigan-atkinsTwodimensionalOverdampedFluctuations2021}. In scattering experiments, short-range or local order manifests as diffuse scattering; a result of static and dynamic deviations from the average structure. Fixed chemical or local structural correlations result in static diffuse scattering, whereas thermal diffuse scattering arises from dynamic displacements due to lattice dynamics \cite{welberryOneHundredYears2016}. Resolving structural correlations in disordered materials has recently become more feasible with the development of high flux, single crystal X-ray and neutron diffuse scattering instruments and sophisticated modeling algorithms \cite{rosenkranzCorelliEfficientSingle2008,yeImplementationCrossCorrelation2018,krogstadReciprocalSpaceImaging2020a, proffenAdvancesTotalScattering2009, simonovYellComputerProgram2014, morganRmcdiscordReverseMonte2021}, opening up enormous opportunities to understand how local order impacts materials properties.

\begin{figure}
\centering
\includegraphics[width=1\linewidth]{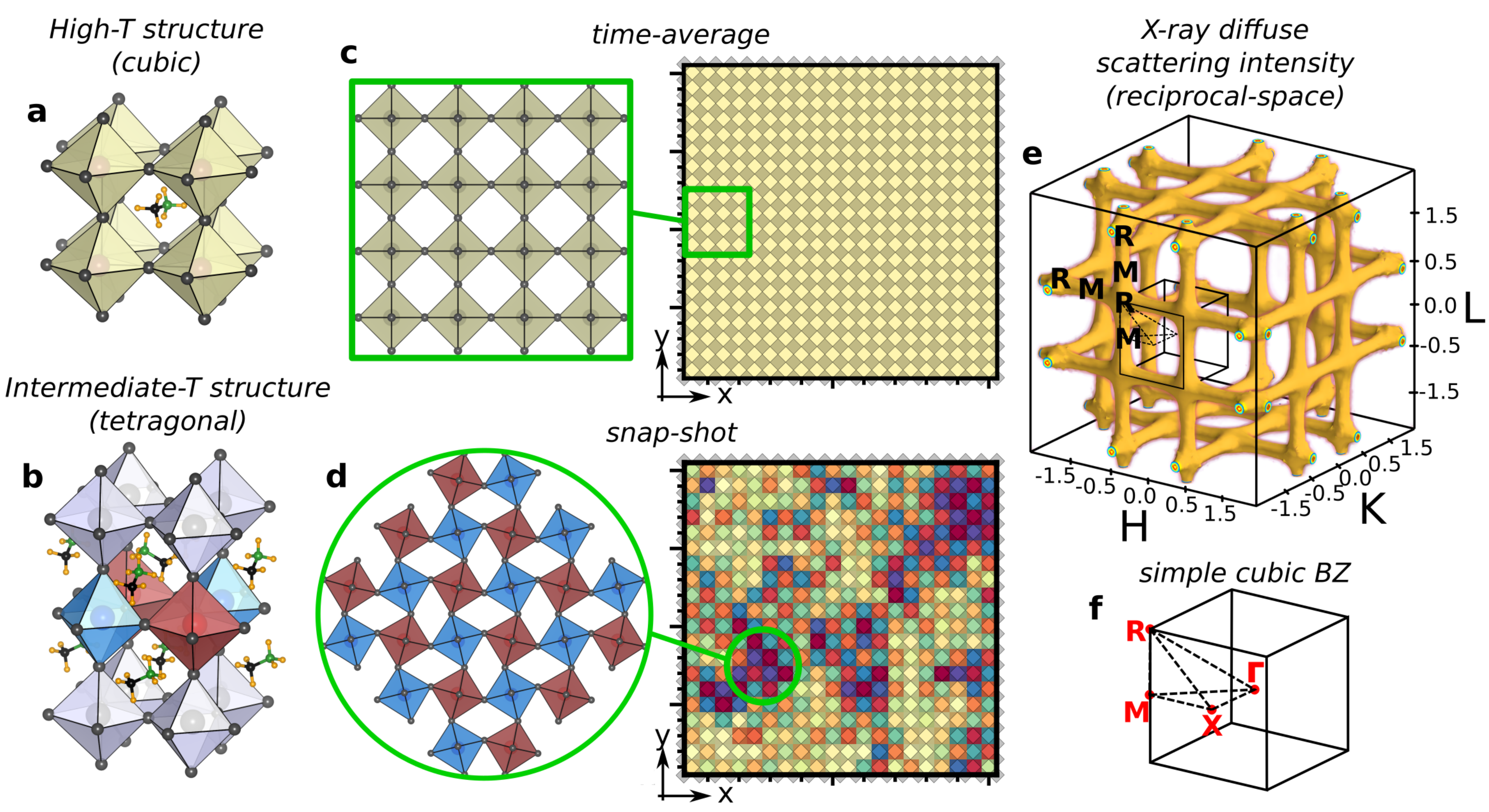}
    \caption{a, c Visualizations of the high-temperature cubic structure and b, the intermediate-temperature tetragonal structure of MAPbI$_3$ and MAPbBr$_3$. MA$^+$ disorder has been omitted for clarity. The cubic-tetragonal transition in MAPbI$_3$ and MAPbBr$_3$ occurs at 328 and 236 K, respectively, by the freezing out of octahedral tilts in an out-of-phase pattern \cite{wellerCompleteStructureCation2015, guoInterplayOrganicCations2017}. d, Instantaneous snapshots of the cubic structure from MD reveal dynamic two-dimensional correlations consisting of tilted PbX$_6$ octahedra that can orient to be perpendicular to any of the three principal axes. e, 3D visualization of the X-ray diffuse scattering intensity generated by the structural correlations. The diffuse volume is indexed by Miller indices H,K,L, of which integer values enumerate each Brillouin Zone (BZ). f, The simple cubic BZ with high symmetry points indicated in red. The diffuse rods in e run along the BZ edge, connecting M and R points.}
    
\label{fig:summary_fig}
\end{figure}

Organic-inorganic metal halide perovskites are a recently re-invigorated class of semiconductors with remarkable optoelectronic performance that defies traditional intuition: Lead-based metal halide perovskites (LHPs) possess a mechanically soft, defect-tolerant crystal lattice with strong structural disorder and mobile ions at modest temperatures \cite{eggerWhatRemainsUnexplained2018}. Fluctuations in the orbital overlaps arising from large thermal displacements of iodide in methylammonium (CH$_3$NH$_3^+$, CD$_3$ND$_3^+$ = MA) lead iodide directly influence the temperature dependence of the electron (or hole) mobility and optical bandgap \cite{mayersHowLatticeCharge2018a}. X-ray and neutron diffraction measurements, which probe long-range order, report that the high-temperature phase is cubic with well-defined Bragg peaks \cite{wellerCompleteStructureCation2015, whitfieldStructuresPhaseTransitions2016, swainsonPhaseTransitionsPerovskite2003}. This cubic perovskite structure is shown in Figure \ref{fig:summary_fig}a and consists of corner-sharing PbX$_6$ (X = I$^-$, Br$^-$) octahedra surrounding a dynamically disordered MA$^+$ cation within the cuboctahedral interstice. Measurements probing short-range order in the high-temperature phase, however, suggest the local structure is of lower symmetry \cite{beecherDirectObservationDynamic2016,cominLatticeDynamicsNature2016,lauritaChemicalTuningDynamic2017, zhaoPolymorphousNatureCubic2020, weadockTestDynamicDomainCritical2020}. The lack of consensus regarding the exact symmetry and dynamics of this enigmatic local structure limits our understanding and control of optoelectronic properties and ion migration in LHPs. Short-range order arising from dynamic two-dimensional correlations of lead bromide octahedra has recently been identified in CsPbBr$_3$ \cite{lanigan-atkinsTwodimensionalOverdampedFluctuations2021}. These correlations are most prominent in the cubic phase above 433 K and may not play a significant role in CsPbBr$_3$-based device operation. The significance and prevalence of such correlations in hybrid LHPs, including contributions from the organic cations, is not resolved. 

Ion migration in LHPs, especially under illumination, is detrimental to device performance and stability yet the origin is not well understood. Consequences of ion migration include formation of space charge potentials at interfaces, accelerated degradation and loss of constituent elements, and light-induced phase segregation in mixed-halide compositions \cite{yuanIonMigrationOrganometal2016}. A complete picture of the high-temperature structure and dynamics is essential to model ion migration pathways accurately \cite{holekevichandrappaCorrelatedOctahedralRotation2021}. It is also important to characterize the short-range order on the organic sublattice, as the rotating MA$^+$ molecular dipoles may screen band-edge charge carriers and extend carrier lifetimes \cite{chenOriginLongLifetime2017}.

We use single crystal X-ray and neutron diffuse scattering and neutron spectroscopy, combined with molecular dynamics (MD) simulations, to uncover the true structure and dynamics of the nominally simple cubic ($Pm\bar{3}m$) phases of MAPbI$_3$ ($>$> 327 K) and MAPbBr$_3$ ($>$> 237 K). We find that the cubic phase (Fig.~\ref{fig:summary_fig}a, c) is comprised of dynamic, two-dimensional roughly circular ``pancakes" of tilted lead halide octahedra which align along any of the three principal axes of the cubic structure and (Fig. ~\ref{fig:summary_fig}b, d) induce additional structural correlations of the organic sublattice in the layers sandwiching the octahedra. Taken together, these regions of dynamic local order are several unit cells in diameter with lifetimes on the order of several picoseconds, resulting in a dynamic landscape for charge carriers and ion migration. This extended dynamic local order was neither observed nor predicted in previous studies of structural dynamics in hybrid LHPs \cite{ferreira_elastic_2018,gold-parkerAcousticPhononLifetimes2018,leguyDynamicsMethylammoniumIons2015,chenRotationalDynamicsOrganic2015}. Embedded within this dynamic structure we find additional static 3D droplets of the intermediate tetragonal phase, consistent with previous reports \cite{weadockTestDynamicDomainCritical2020, cominLatticeDynamicsNature2016}. The structural correlations on the MA$^+$ sublattice are a unique characteristic in these hybrid systems with implications discussed below. 

\begin{figure}[h!]
    \centering
    \includegraphics[width = 1\linewidth]{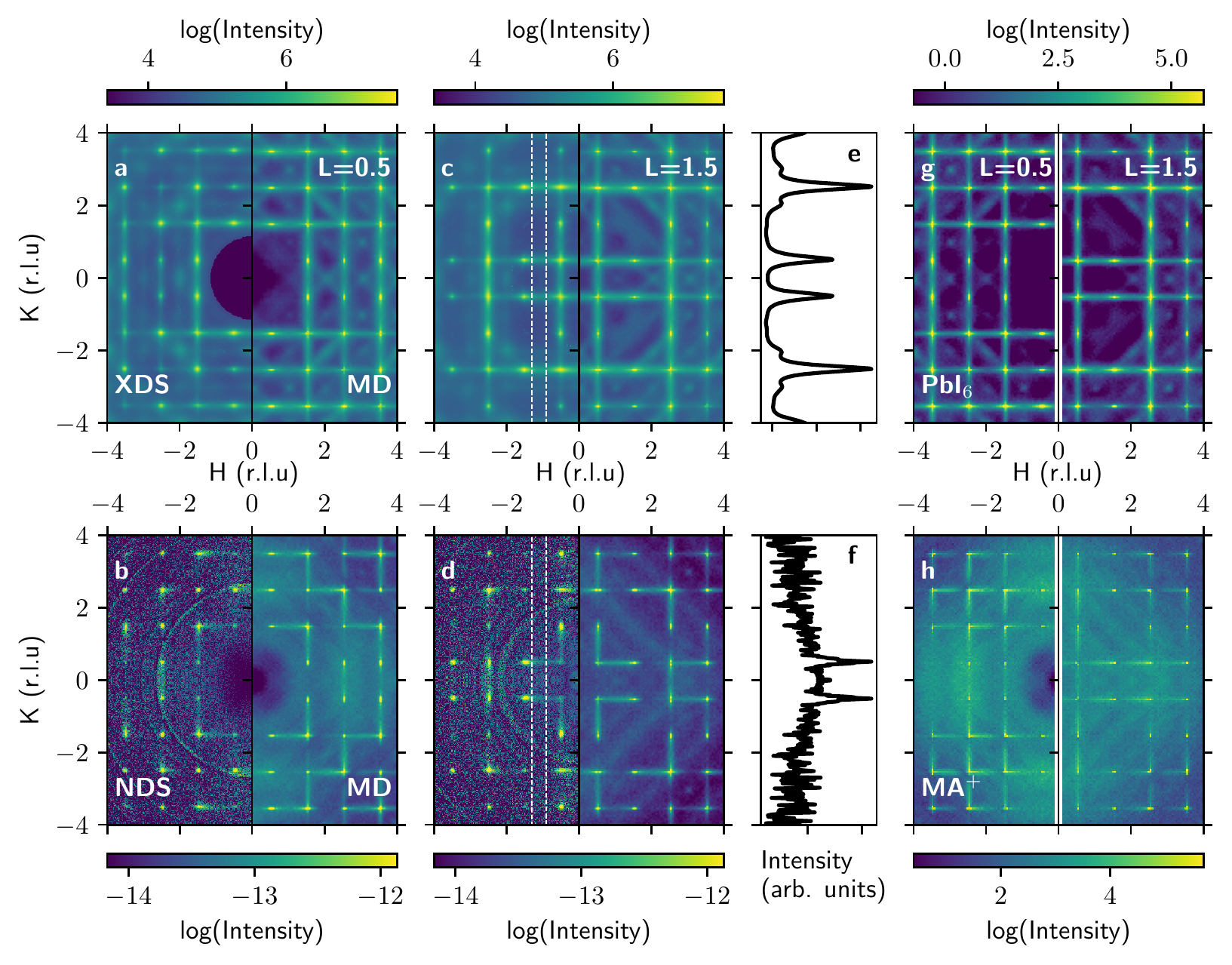}
    \caption{Experimental $S(\mathbf{Q})$ for MAPbI$_3$ measured with XDS (a,c left panels) and NDS (b,d left panels) at 345 K. The XDS data are inherently energy integrated whereas the NDS data are static with a 0.9 meV FWHM energy bandwidth. Panels a,b correspond to the L = 0.5 plane, and panels c,d the L = 1.5 plane. No Bragg intensity is expected in these half-order planes; therefore any intensity is diffuse scattering. The faint circular rings in the experimental NDS data are residual Debye-Scherrer diffraction rings from the aluminum sample holder. Theoretical $S(\mathbf{Q})$, calculated from MD simulations ($\pm$1 meV bandwith), are plotted in the right panels of a-d. In a,c, X-ray form factors were used whereas b,d,g,h were calculated using neutron scattering lengths (see supplementary information \cite{supp_info}). e,f plot one-dimensional linecuts along K, integrating 0.7:H:1.3 and 1.4:L:1.6, which highlight the differences between XDS and NDS. Panels g,h plot the PbI$_6$ and MA$^+$ contributions to the calculated neutron $S(\mathbf{Q})$, respectively, with the left panels corresponding to L = 0.5 and the right panels L = 1.5. We note that the NDS plotted here corresponds to the elastic scattering channel of the CORELLI spectrometer ($\sim 0.9$ meV full width at half-maximum for $E_{\mathrm{i}} = 30$ meV)\cite{yeImplementationCrossCorrelation2018}.}
    \label{fig:fig2} 
\end{figure}

The experimental scattering function $S(\mathbf{Q})$ from the cubic phase of MAPbI$_3$ at 345 K is shown in Fig.~\ref{fig:fig2}. Specifically, we present the L = 0.5, 1.5 planes measured with both X-ray (XDS, Fig.~\ref{fig:fig2}a,c, left panels) and neutron diffuse scattering (NDS, Fig~\ref{fig:fig2}b,d, left panels). $S(\mathbf{Q})$ calculated from atomic trajectories obtained with MD simulations of MAPbI$_3$, see Methods, is plotted in the right-hand panels of a-d and both panels of g,h for comparison.

The experimental $S(\mathbf{Q})$ of the cubic phase of MAPbBr$_3$ at 250 K is plotted in Fig. S1 and show diffuse scattering intensity profiles for XDS and NDS that are nearly identical to MAPbI$_3$.

Our approach to solving the true structure of the nominally simple cubic MAPbI$_3$ and MAPbBr$_3$ involves decomposing the observed diffuse scattering profile into four components and analyzing their energy and \textbf{Q}-dependence. These components include: (1) rods of constant (XDS) or varied (NDS) intensity spanning the BZ edge (M-R direction in Fig.~\ref{fig:summary_fig}, $q = [0.5, 0.5, L]$); (2) an additional contribution centered at the R-point [$q =(0.5, 0.5, 0.5)$]; (3) broad intensity centered at the X-point [$q =(0.5, 0, 0)$] observed previously and discussed in the SI; and (4) contributions from the MA$^+$ sublattice deduced from the alternating diffuse rod intensity profile observed in NDS but not XDS. This difference between XDS and NDS is highlighted with one-dimensional line cuts of XDS and NDS data shown in Figs.~\ref{fig:fig2}e and f. Specifically, the peaks at K = $\pm2.5$ originating from the extended rod along H in XDS are not observed in the NDS data. Our analysis is complemented by MD simulations of MAPbI$_3$ which reproduce the experimental diffuse scattering. The experimental and calculated XDS (Fig.~\ref{fig:fig2}) show remarkable agreement. Comparing the two results in a root-mean-square error of 3.9\% (see Figs.~S2,3 and related discussion) after adjusting only the background and scaling of the calculated $S(\mathbf{Q})$. The agreement between the experimental and calculated NDS is very good, however there are small differences in intensities highlighted in Fig.~S4 that likely arise from the classical potentials used here. Given the good agreement, we examine the real-space structure simulated with MD to determine the origin of the diffuse scattering.

\begin{figure}
\centering
\includegraphics[width=0.75\linewidth]{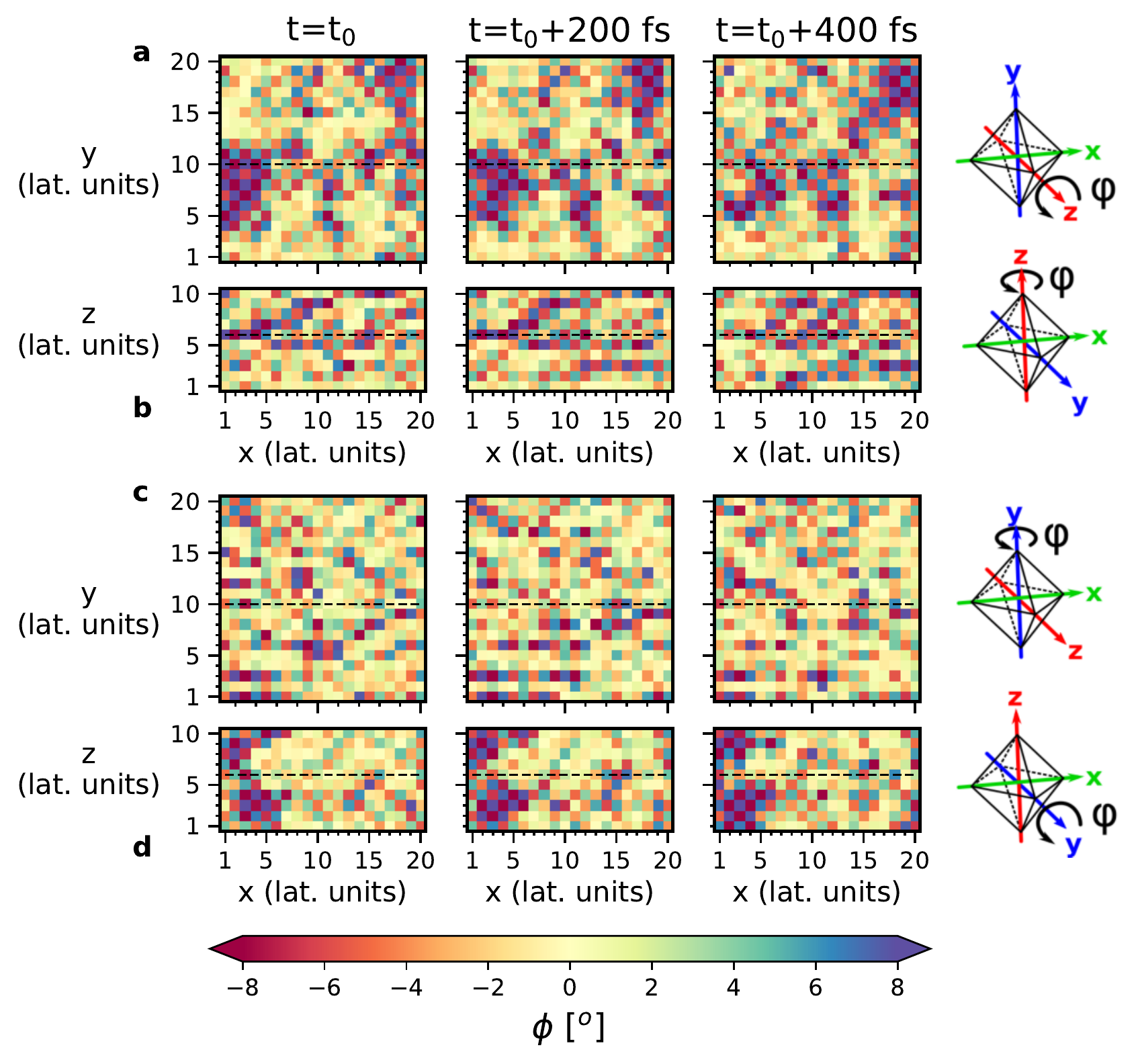}
    \caption{Spatiotemporal dynamics of PbI$_6$ octahedral rotations in cubic MAPbI$_3$ calculated from MD. Each pixel encodes the octahedral rotation angle $\phi$ about the $z$-axis [a and b] and the $y$-axis [c and d]. The columns are offset in time by $\Delta$t=200 fs to highlight the dynamic nature. The top rows [a and c] are in the $x$-$y$ plane and the bottom rows [b and d] are in the $x$-$z$ plane. In a and d, the rotation is about the axis perpendicular to the plane as shown by the diagrams on the right. In b and c, the rotation is about an axis parallel to the figure's vertical axis. The dashed lines indicate the intersection between the $x$-$y$ plane in a and c and the $x$-$z$ plane in b and d. Note that the two-dimensional regions visible in a, d are one unit cell thick in the normal direction b, c.}
\label{fig:snapshots}
\end{figure}

The XDS rods of diffuse scattering have constant intensity along the entire BZ edge (Fig.~S5) and a width larger than the instrument resolution ($\approx 0.02 \AA^{-1}$). This lineshape arises from two-dimensional structural correlations in real-space. We show these correlations in Fig.~\ref{fig:snapshots}, which presents a simplified visualization of the MD simulation. These plots track PbI$_6$ octahedral rotations, defined by azimuthal rotation angle $\phi$ about the [001] (Fig.~\ref{fig:snapshots}a,b) or [010] (Fig.~\ref{fig:snapshots}c,d) directions. Correlated regions of alternating large tilts are identified by neighboring red and blue pixels (representing octahedra) in Fig.~\ref{fig:snapshots}a and d. Along the axis of rotation indicated in the corresponding schematic the tilts are uncorrelated between planes (Fig.~\ref{fig:snapshots}b,c), revealing the existence of two-dimensional pancakes of tilted octahedra with a diameter on the order of 5 unit cells. These regions are reminiscent of tilts associated with the tetragonal structure in Fig.~\ref{fig:summary_fig}b. The tilts generate in-plane antiphase structural correlations with a periodicity of $2a$. The associated wave vector of the correlation is $q=2\pi / 2a =\pi/a$, which is the zone boundary for the simple cubic BZ. The tilted regions are confined to a single PbI$_6$ sheet and therefore the diffuse scattering intensity manifests as rods. 

The same structural correlations are present in the cubic phase of MAPbBr$_3$, since the XDS and NDS are qualitatively identical to that of MAPbI$_3$ (Fig.~S1, \ref{fig:fig2}). Extended diffuse rods observed in the all-inorganic CsPbBr$_3$ \cite{lanigan-atkinsTwodimensionalOverdampedFluctuations2021} point to the existence of two-dimensional structural correlations of lead halide octahedra in all LHPs with a high temperature cubic structure.

Intermolecular structural correlations in the MA$^+$ sublattice are implied from the alternating intensity pattern of the diffuse rods present in NDS but not XDS. The MA$^+$ correlations are explored in the calculated $S(\mathbf{Q})$ in Figs.~\ref{fig:fig2}g,h, and S6-8. Specifically, we set the neutron scattering lengths of Pb and I to zero in the calculation to isolate the contribution from the MA$^+$ cation (Figs.~\ref{fig:fig2}h, S6c, g, k and S7), and  C, N, and D to zero to isolate contributions from the inorganic octahedra (Figs.~\ref{fig:fig2}g, S6b, f, j and S7). Diffuse scattering from the inorganic framework manifests as extended diffuse rods along the zone edge, resembling the experimental XDS as the X-ray atomic form factors for C, N, D are small compared to those for Pb and I and contribute little intensity. The calculated $S(\mathbf{Q})$ from CD$_3$ND$_3^+$ shows remarkable behavior (Figs.~\ref{fig:fig2}h, S6, S7), namely well-defined diffuse rods along the zone edges in addition to broad, isotropic intensity. This broad, isotropic component is a result of uncorrelated MA$^+$ dynamics \cite{leguyDynamicsMethylammoniumIons2015, chenRotationalDynamicsOrganic2015} commonly associated with the nominally cubic phases of LHPs. The presence of additional zone edge intensity, however, shows that previously unreported intermolecular structural correlations exist on the MA$^+$ sublattice. These local correlations are not purely 2D as the intensity varies along individual diffuse rods in the MA$^+$ $S(\mathbf{Q})$ (Fig.~S6,7). Layers of PbX$_6$ octahedra are sandwiched between layers of MA$^+$ cations, therefore we expect the out-of-plane MA$^+$ structural correlations to extend at least two unit cells. Finally, the MA$^+$ and PbX$_6$ correlations are connected as evidenced by a nonzero interference term, shown in Fig.~S8, which results in the alternating diffuse intensity profile observed in NDS.  

We propose that MA$^+$ orientation is driven by PbX$_6$ tilts; the MA$^+$ molecules orient to favor the lowest energy configuration defined by the distorted cuboctahedral geometry of the tilted regions and electrostatic interactions between lead halide octahedra and polar MA$^+$ molecules \cite{zhu2019mixed,lahnsteiner2016room}. The existence of 2D correlations on the PbBr$_6$ sublattice in CsPbBr$_3$ supports this prediction as the Cs$^+$ cations are not expected to influence the PbBr$_6$ octahedra through hydrogen bonding. \cite{lanigan-atkinsTwodimensionalOverdampedFluctuations2021}.

The lateral size of 2D pancakes, given by the correlation length, $\xi$, is obtained from the $\mathbf{Q}$-linewidth of the XDS and NDS diffuse rods. We fit the intensity of one-dimensional cuts in $\mathrm{\AA}^{-1}$ across various diffuse rods to a Lorentzian lineshape in accordance with Ornstein-Zernike theory for exponentially decaying spatial correlations:
\begin{equation}
    S(\mathbf{q}) \propto \frac{\Gamma}{\Gamma^2 + (q-q_0)^2}
\label{ozt}
\end{equation}

where $\Gamma = 1/\xi$ is the half-width at half-maximum (HWHM) of the fitted Lorentzian and $\xi$ is the correlation length \cite{xuThreedimensionalMappingDiffuse2004, valeCriticalFluctuationsSpin2019}. Correlation lengths of 4-6 unit cells ($\sim$3 nm) in diameter are obtained for MAPbBr$_3$ and MAPbI$_3$ as reported in Table~S1. Similar $\xi$ are obtained from NDS and XDS, indicating the in-plane $\xi$ of the 2D pancakes are nearly equivalent for the PbX$_6$ and MA$^+$ sublattices. We also calculate $\xi$ directly from the MD simulations, see Figs. S9-12 and related discussion, and find values consistent with experiment.

\begin{figure}
\centering
\includegraphics[width=1\linewidth]{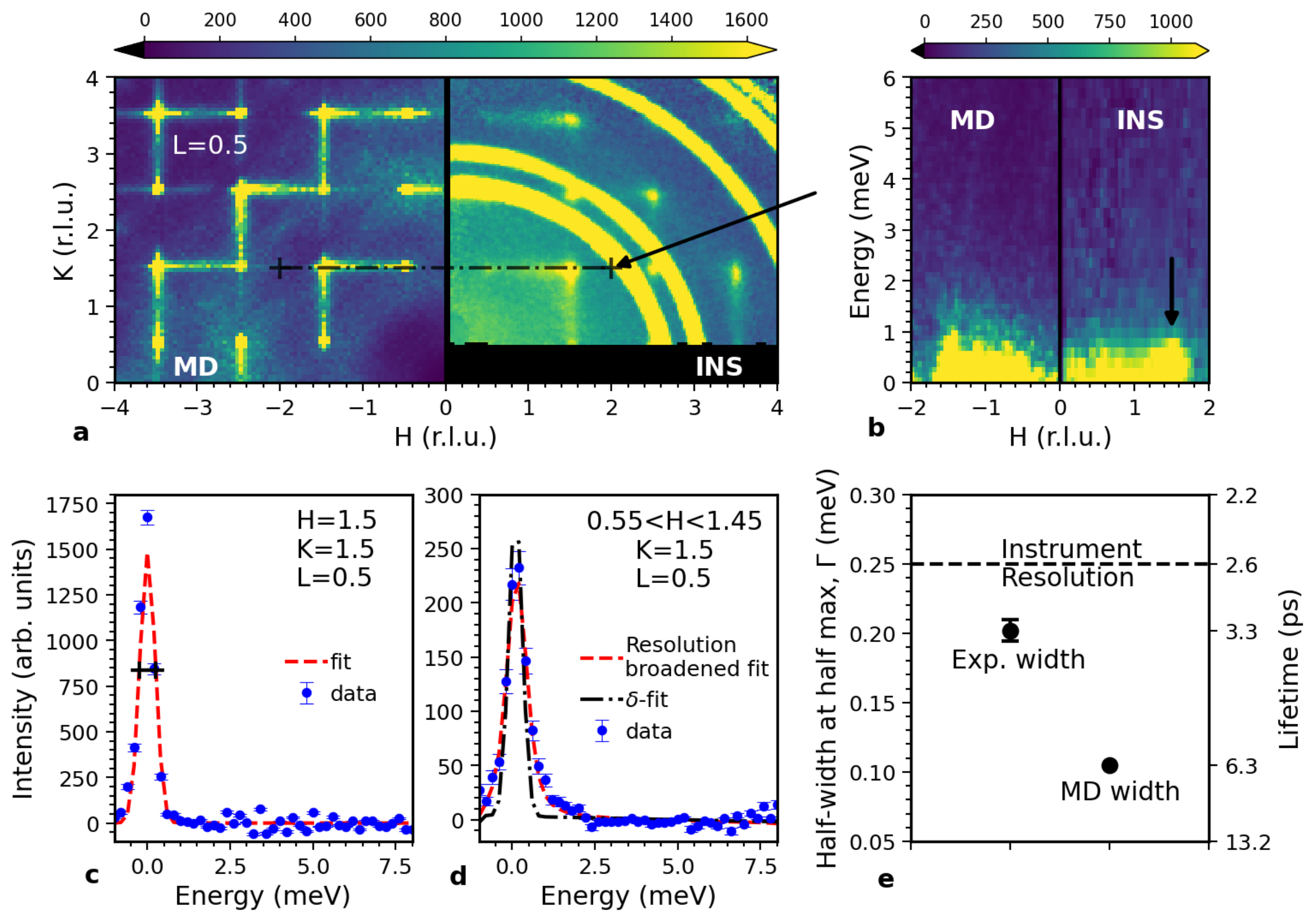}
    \caption{Theoretical and experimental $S(\mathbf{Q}, E)$ of MAPbI$_3$, highlighting the energy dependence of the diffuse scattering rods. Neutron inelastic scattering data (right panels of a,b; and c,d) were measured at 340 K. In a, the $S(\mathbf{Q})$ from MD and experiment is integrated from $-1 \leq E \leq 1$meV. Powder rings from the aluminum sample holder are visible in a. Constant-\textbf{Q} scans at the R-point (c, integrated along 1.45:H:1.55, 1.45:K:1.55, 0.45:L:0.55) and along the diffuse rod (d, integrated along 0.55:H:1.45, 1.45:K:1.55, 0.45:L:0.55) are fit to Gaussian and resolution-broadened Lorentzian lineshapes with resolution linewidth of 0.5 meV FHWM, respectively (red dashed lines). The black line in c indicates the 0.5 meV resolution FWHM. In d, we include the resolution lineshape (black dash-dot line), demonstrating a finite width beyond the resolution. In e, the average HWHM (and corresponding lifetime) obtained from Lorentzian fits to constant-\textbf{Q} scans are plotted for experiments and MD. The dashed line represents the energy resolution of the MERLIN spectrometer.}
\label{fig:fig3}
\end{figure}

Molecular dynamics simulations show that the 2D structural correlations are transient and diffusive, with dynamics of the octahedral correlations tracked in Fig.~\ref{fig:snapshots} and Supplementary Videos 1 and 2. We investigate the dynamics of the PbI$_6$ and MA$^+$ correlations by evaluating the energy dependence obtained from the dynamical structure factor $S(\mathbf{Q}, E =\hbar\omega)$ measured with neutron inelastic spectroscopy (INS) on MAPbI$_3$ at 340 K. The $S(\mathbf{Q}, E)$ in the L = 0.5 plane, integrated from $-1\leq E\leq 1$ meV is shown in the right-hand panel of Fig.~\ref{fig:fig3}a and matches that observed with NDS (Fig.~\ref{fig:fig2}b) and that calculated from MD (left-hand panel of Fig.~\ref{fig:fig3}b). A slice of $S(\mathbf{Q}, E)$ along the diffuse rod presented in Fig.~\ref{fig:fig3}b shows that the intensity is centered at $E = 0$ meV with no inelastic component visible. Incoherent-background-subtracted energy scans (integrated along the length of the diffuse rod) show a finite, quasielastic component as exemplified in Fig.~\ref{fig:fig3}d. Quasielastic scattering is a result of diffusive or relaxational motions involving small energy transfers, manifesting as a peak centered at $E = 0$ meV with a non-zero linewidth. The scattering is fit to a relaxational (Lorentzian) model, broadened by the resolution function, as shown in Fig.~\ref{fig:fig3}d. The average energy HWHM, plotted in Fig.~\ref{fig:fig3}e, is $0.20\pm0.01$ meV with a corresponding lifetime $\tau = \hbar/\mathrm{HWHM}$ of $3.25\pm0.13$ ps. Lifetimes are obtained in the same manner from the calculated $S(\mathbf{Q}, E)$ as shown in Fig.~S13, and the average lifetime of 6.3 ps is reported in Fig.~\ref{fig:fig3}e. We note that these lifetimes are consistent with the mean residence time for rotations of the C-N bond in cubic MAPbI$_3$, suggesting that MA$^+$ reorientations are in part driven by the PbX$_6$ correlations \cite{chenRotationalDynamicsOrganic2015}. Overall, the structural correlations are dynamic with finite lifetime of several picoseconds. 

R-point scattering ($\mathbf{Q} = [1.5,1.5,0.5]$, Fig.~\ref{fig:fig3}c) noted above has been investigated previously and is attributed to static droplets of the intermediate-temperature tetragonal phase embedded within the high-temperature cubic phase, possibly nucleating about defects \cite{weadockTestDynamicDomainCritical2020}.  We observe no inelastic scattering at the R-points in both the INS or MD $S(\mathbf{Q} = (H, 1.5, 0.5) ,E)$ in Fig.~\ref{fig:fig3}b. Incoherent-background-subtracted energy scans at a constant \textbf{Q} corresponding to R-points are best fit by a Gaussian lineshape with resolution limited linewidth (Fig.~\ref{fig:fig3}c), hence this scattering is static with $\xi =2$ nm (Table~S1), consistent with previous reports \cite{weadockTestDynamicDomainCritical2020}. These droplets are distinct from the 2D pancakes but we are not able to resolve the spatial distribution of these two components. We expect that both structures influence MA$^+$ orientation. 

The appearance of small regions of a lower temperature phase above the phase transition temperature is often a hallmark of critical scattering. For phase transitions with critical fluctuations, there is a marked jump in intensity of the associated low temperature phase as the system is cooled through the transition \cite{stirling_critical_1996}. We investigate the temperature dependence of the diffuse scattering in Figs.~S13,14 through the cubic-tetragonal transition temperatures for MAPbI$_3$ and MAPbBr$_3$. The intensity of the diffuse rods increases continuously with decreasing temperature through the transition, however no significant and abrupt jump in intensity is observed. Furthermore, the two-dimensional nature of the dynamic structural correlations is not typical of critical behavior at phase transitions. For the R-point scattering, however, there is a large jump in intensity as the cubic R-point transforms to the $\Gamma$-point of the tetragonal phase, giving Bragg scattering. The critical nature of the R-point scattering has been addressed previously \cite{weadockTestDynamicDomainCritical2020, cominLatticeDynamicsNature2016}.

Our results show that the simple cubic phase of MAPbI$_3$ and MAPbBr$_3$ is in fact a composite structure containing dynamic 2D structural correlations and static tetragonal droplets with $\xi$ of a few nanometers. MA$^+$ correlations, induced by octahedral tilts, extend 2-3 unit cells in the normal direction. The simple cubic phase is only recovered when this composite is averaged over space and time (Fig. \ref{fig:summary_fig} a-d). Structural fluctuations of lead halide octahedra generate large variations in the optical bandgap and electron-phonon coupling \cite{lanigan-atkinsTwodimensionalOverdampedFluctuations2021, mayersHowLatticeCharge2018a, zhaoPolymorphousNatureCubic2020}, therefore it is essential to incorporate the observed correlations when modeling optoelectronic properties. We expect the dynamic correlations and static droplets to introduce spatial anisotropy to the electronic potential landscape with a longer lifetime than variations resulting from uncorrelated dynamic disorder. Furthermore, we must consider the effect of local correlations of the organic sublattice on LHP properties. MA$^+$ has a large electric dipole moment of 2.3 Debye along the C-N axis \cite{chenOriginLongLifetime2017}. Uncorrelated dynamic disorder of the MA$^+$ sublattice, as previously reported, would generate a zero or negligible dipole moment. The MA$^+$ correlations observed here, however, may generate a net dipole moment resulting in transient, local ferroelectric or antiferroelectric domains with lifetimes on the order of 3-6 ps. These domains influence electron-hole recombination and shift band alignment \cite{biEnhancedPhotovoltaicProperties2017, liuChemicalNatureFerroelastic2018}. The dynamic structural correlations also affect ion migration. In solid-state ion conductors, rotational dynamics of polyanions are known to strongly impact cation mobility  \cite{smithLowtemperaturePaddlewheelEffect2020}. Since the lifetime of the dynamic structural correlations (3-6 ps) exceeds the calculated time between attempted halide jumps (1 ps) \cite{frostWhatMovingHybrid2016}, a prefactor in calculating diffusivity, the correlations appear static and must be considered in future calculations. These correlations may impose a barrier to halide migration and reduce diffusivity \cite{holekevichandrappaCorrelatedOctahedralRotation2021}.

The two-dimensional dynamic structural correlations we observe may arise from the strongly anharmonic lattice dynamics observed in LHPs \cite{zhu2019mixed, gold-parkerAcousticPhononLifetimes2018, songvilayCommonAcousticPhonon2019, ferreira_elastic_2018}. Indeed, the structural correlations in CsPbBr$_3$ are reportedly driven by soft, overdamped, anharmonic phonons at the BZ edge \cite{lanigan-atkinsTwodimensionalOverdampedFluctuations2021}. We do not observe dispersive phonons along the BZ edge M-R branch in Fig.~\ref{fig:fig3}b, therefore these modes may be overdamped for MAPbI$_3$ and MAPbBr$_3$ as well. The structural correlations we observe on the PbX$_6$ sublattice have a longer lifetime than those in CsPbBr$_3$, and are prevalent at device relevant temperatures. The additional intermolecular correlations between MA$^+$ are not present in CsPbBr$_3$ and contribute to the improved optoelectronic properties of hybrid LHPs.

\section{Acknowledgements}

The authors acknowledge helpful discussions with Xixi Qin, Volker Blum, and Alex Zunger. This work (X-ray and neutron scattering) was supported by the Center for Hybrid Organic Inorganic Semiconductors for Energy, and Energy Frontier Research Center funded by the Office of Basic Energy Sciences, an office of science within the US Department of Energy (DOE). J. A. V. acknowledges fellowship support from the Stanford University Office of the Vice Provost of Graduate Education and the National Science Foundation Graduate Research Fellowship Program under Grant No. DGE – 1656518. H. I. K. acknowledges funding through the DOE Office of Basic Energy Sciences, Division of Materials Science and Engineering, under Contract No. DE-AC02-76SF0051. T.C.S. and D.R. acknowledge funding by the DOE Office of Basic Energy Sciences, Office of Science, under Contract No. DE-SC0006939. A portion of this research used resources at the Spallation Neutron Source, a DOE Office of Science User Facility operated by the Oak Ridge National Laboratory. Use of the Advanced Photon Source at Argonne National Laboratory was supported by the U.S. Department of Energy, Office of Science, Office of Basic Energy Sciences, under Contract No. DE-AC02-06CH11357. Experiments at the ISIS Pulsed Neutron and Muon Source were supported by a beamtime allocation from the Science and Technology Facilities Council. Any mention of commercial products here is for information only; it does not imply recommendation or endorsement by the National Institute of Standards and Technology.

\section{Methods}\label{sec:methods}

\subsection{Single crystal growth}\label{singlextal}

\subsubsection{Materials}
Methyl-d$_3$-amine (CD$_3$NH$_2$, 99 at. \% D) was purchased from Sigma-Aldrich. Hydrobromic acid (concentrated HBr solution, 48 \% in water) was purchased from Acros Organics. Hydroiodic acid (concentrated HI solution, $\geq$47 \% in water with $\leq$1.5 \% hypophosphorous acid) was purchased from Sigma-Aldrich. Methanol-d$_4$ and ethanol-d$_6$ were purchased from Acros Organics and Cambridge Isotope Laboratories, Inc., respectively, with $\geq$99 at. \% D purity. PbI$_2$ (99.999+ \%-Pb) and PbBr$_2$ (98+ \%, extra pure) were purchased from Strem Chemicals, Inc. and Acros Organics, respectively. CH$_3$NH$_3$I and CH$_3$NH$_3$Br were purchased from Greatcell Solar Materials. All solvents were of reagent grade or higher purity and anhydrous solvents were stored over molecular sieves or used directly from a JC Meyer solvent purification system.

\subsubsection{Synthesis of CD$_3$ND$_3$X$_3$ (X = Br, I)}
Methyl-d$_3$-amine was bubbled through ethanol (in a round bottom flask with flowing N$_2$ in the headspace) in an ice water bath under vigorous stirring for approximately 2 minutes. Excess concentrated HBr was subsequently added dropwise with stirring. After the solution warmed naturally to room temperature (RT), the solvent was removed under reduced pressure at 60 \textdegree C. The resulting oil was re-dissolved in anhydrous ethanol and the product, CD$_3$NH$_3$Br, was precipitated as a colorless solid upon addition of the solution to cold diethyl ether. The solid was isolated by vacuum filtration and subsequently re-dispersed, sonicated, and isolated from diethyl ether three times before drying the product overnight under reduced pressure.

Solid CD$_3$NH$_3$Br was dissolved in ethanol-d$_6$ and stirred for 3 h at 60 \textdegree C before precipitating the product in anhydrous diethyl ether and removing excess solvent under reduced pressure. This hydrogen-deuterium exchange was repeated twice more before finally isolating and drying the product, CD$_3$ND$_3$Br, at 65 \textdegree C overnight.

The synthesis of CD$_3$ND$_3$I followed an analogous procedure as for CD$_3$ND$_3$Br, except for the substitutions of: ethanol (or ethanol-d$_6$ during the exchange) for methanol (or methanol-d$_4$), and concentrated HBr for concentrated HI.

\subsubsection{Crystallization of CD$_3$ND$_3$PbBr$_3$}
Solid CD$_3$ND$_3$Br (0.235 g, 2 mmol) and PbBr$_2$ (0.731 g, 2 mmol) were combined in 2 mL of anhydrous dimethylformamide and the mixture was sonicated for 20 minutes. Once dissolved, the solution was passed through a 0.22-$\mu$m hydrophilic PTFE filter and transferred to an oil bath at 78 \textdegree C. Crystals of CD$_3$ND$_3$PbBr$_3$ slowly nucleated at the interface between the solution and the surface of the vial as the temperature increased slowly (1-3 \textdegree C h$^{-1}$) from 78 \textdegree C (up to  85-100 \textdegree C). Crystals were isolated from the hot solution and the residual precursor solution was quickly removed from the surface of the crystal. Large crystals ($\geq$200 mg) were obtained by isolating small crystals (25-50 mg) as seeds and re-filtering the precursor solution into a fresh vial with one seed crystal and quickly transferring back to the oil bath at 78 \textdegree C. Repeating the temperature ramp described above resulted in large crystal growth after a period of minor re-dissolution from the seed. 

\subsubsection{Crystallization of CH$_3$NH$_3$PbBr$_3$}
A 1-M solution of CH$_3$NH$_3$Br and PbBr$_2$ in anhydrous dimethylformamide was prepared and sonicated for 1 h. Then 1 mL of the precursor solution was passed through a glass microfiber filter and transferred to a shell vial. The shell vial was placed inside a larger vial containing 4 mL of dichloromethane, the anti-solvent for the vapor diffusion crystallization, and sealed. Millimeter-scale (CH$_3$NH$_3$)PbBr$_3$ crystals formed over the course of hours to days at room temperature. 

\subsubsection{Crystallization of CD$_3$ND$_3$PbI$_3$ and CH$_3$NH$_3$PbI$_3$}
Solid CD$_3$ND$_3$I (0.594 g, 3.6 mmol) and PbI$_2$ (1.66 g, 3.6 mmol) were combined in 3 mL of anhydrous $\gamma$-butyrolactone and dissolved at 70 \textdegree C with stirring. The solution was hot-filtered through a through a 0.22-$\mu$m PTFE filter and immediately transferred to an oil bath at 70 \textdegree C. The temperature was increased to 110 \textdegree C over the course of approximately 2 h, followed by a slow ramp (1-3 \textdegree C h$^{-1}$) up to 120-130 \textdegree C. Crystals were isolated and re-seeded as described for CD$_3$ND$_3$PbBr$_3$ to reach the desired size.

The crystallization of CH$_3$NH$_3$PbI$_3$ followed an analogous procedure as for CD$_3$ND$_3$PbI$_3$, except for the substitution of CD$_3$ND$_3$I for CH$_3$NH$_3$I. 

\subsection{Diffuse X-ray scattering}\label{xds}
Reciprocal space maps, including Bragg and diffuse scattering contributions, were collected in transmission geometry at the Advanced Photon Source Sector 6-ID-D using monochromatic 86.9 keV X-rays. This X-ray scattering is inherently energy integrated, incorporating contributions from both static and thermal diffuse disorder. Single crystal samples of (CD$_3$ND$_3$)PbI$_3$, (CH$_3$NH$_3$)PbI$_3$, and (CH$_3$NH$_3$)PbBr$_3$, $\sim 500 \mu$m in size were mounted on the tip of Kapton capillaries. At each temperature, the sample was rotated $365^{\circ}$ at $1^{\circ} \mathrm{s}^{-1}$ with images collected every 0.1 s on a Pilatus 2M CdTe area detector. Sample temperatures were varied between 150 - 300 K with an Oxford N-Helix (check) Cryostream and 300 - 360 K with a hot nitrogen blower. The raw images are first processed with a peak finding algorithm to determine and refine an orientation matrix, and then rebinned into a reciprocal space volume $\pm$10 r.l.u. a side using the CCTW reduction workflow\cite{krogstadReciprocalSpaceImaging2020a,Cctw}. The dataset is symmetrized using cubic point group operations to remove missing data resulting from gaps between detector pixel banks. 

Sample damage is evident from changes in the scattering intensity with prolonged exposure. The damage thresholds were determined by identifying the onset of additional scattering intensity in subsequent measurements of the same sample at constant temperature. In MAPbI$_3$, damage is apparent after 90 minutes of measurement time (including time when the shutter is off in between each rotation for data collection), therefore all diffuse scattering analyzed here was collected on a fresh sample within 90 minutes.

\subsection{Diffuse neutron scattering}\label{nds}

Neutron diffuse scattering measurements were performed on the CORELLI spectrometer at the Spallation Neutron Source at Oak Ridge National Lab in Oak Ridge, Tennessee, USA\cite{rosenkranzCorelliEfficientSingle2008,yeImplementationCrossCorrelation2018}. Deuterated MAPbBr$_3$ (300, 350 mg, rectangular prism) and MAPbI$_3$ (220 mg, rhombic dodecahedron) crystals were used to reduce incoherent background contributions from hydrogen. The crystals were mounted in a CCR cryostat and oriented in the (HK0) scattering plane. The incident neutron energies range from 10 - 200 meV, and the 2$\theta$ coverage spans -30 to 145$^{\circ}$. MAPbBr$_3$ scattering data was collected from 150 - 300 K for 4-6 hours per temperature, and MAPbI$_3$ scattering data was collected from 300 - 400 K at 12 hours per temperature to account for smaller sample size and increased background from the thermal shielding. Diffuse scattering volumes $\pm$6 r.l.u. a side were obtained by reducing the raw data using defined workflows implemented in \texttt{MANTID}\cite{arnoldMantidDataAnalysis2014}. The contributions from inelastic scattering are removed by implementing the cross-correlation technique in the \texttt{MANTID} reduction workflow. The cross-correlation chopper selects elastically scattered neutrons within a resolution bandwidth of 0.9 meV full-width at half maximum \cite{yeImplementationCrossCorrelation2018}. Background artefacts including the aluminum sample enclosure are removed by subtracting the highest temperature datasets (300 K for MAPbBr$_3$, 400 K for MAPbI$_3$), normalized by the Bragg peaks, where little to no diffuse scattering is detected.

\subsection{Neutron inelastic scattering}\label{ins}

Neutron inelastic scattering measurements were performed on the Merlin direct geometry chopper spectrometer at the ISIS Neutron and Muon source at the Rutherford Appleton Laboratory in Didcot, UK\cite{bewleyMERLINNewHigh2006}. A deuterated single crystal of MAPbI$_3$ weighing 698 mg was oriented in the (HKK) scattering plane and mounted in a Brookhaven-style aluminum sample can. Temperature control was performed using a closed-cycle refrigerator with heaters attached directly to the sample can. Three incident energies of 11, 22, and 65 meV were selected by utilizing the repetition rate multiplication mode with the Fermi chopper set to 250 Hz. The $S(\mathbf{Q}, E)$ was collected at 340 K by rotating the sample 120$^{\circ}$ in 0.5$^{\circ}$ steps, and a radial collimator was used to reduce scattering from the aluminum sample can. Data reduction was performed using the \texttt{HORACE} data reduction suite and analyzed with Phonon Explorer and the National Institute of Standards and Technology (NIST) Center for Neutron Research (NCNR) \texttt{Data Analysis and Visualization Environment}  (DAVE) software packages \cite{ewingsHoraceSoftwareAnalysis2016a, reznikAutomatingAnalysisNeutron2020,phonon-explorer, azuahDAVEComprehensiveSoftware2009}.

Energy scans for linewidth analysis were sliced from the 11 meV incident energy $S(\mathbf{Q}, E)$ dataset. The resolution linewidth of this dataset was determined from constant-$\mathbf{Q}$ scans at $\mathbf{Q}$ values unique to the aluminum Debye-Scherrer rings. Several constant-$\mathbf{Q}$ scans obtained in this manner were fit to a Gaussian lineshape with constant background to obtain an average resolution linewidth of 0.5 meV FWHM. A representative constant-$\mathbf{Q}$ scan was used as the resolution function in subsequent analyses. The incoherent background was determined from constant-$\mathbf{Q}$ scans at $\mathbf{Q}$ values which contain only incoherent scattering with no contributions from diffuse rods, Bragg peaks, or aluminum Debye-Scherrer rings. Several background scans were averaged, then subtracted from the constant-$\mathbf{Q}$ scans along the diffuse rods and at the R-point to remove the incoherent background contribution and isolate the dynamics associated with these features.

Additional high-resolution neutron spectroscopy was performed on the cold neutron triple-axis spectrometer (SPINS) at the NCNR in Gaithersburg, MD, USA. The same deuterated single crystal of MAPbI$_3$ weighing 698 mg was oriented in the (HKK) scattering plane and mounted in a Brookhaven-style aluminum sample can. Constant-$\mathbf{Q}$ scans, in which we vary the energy transfer while maintaining constant momentum transfer $\mathbf{Q}$, were performed with a fixed final energy $E_\mathrm{F} = 5$ meV. We compare constant-$\mathbf{Q}$ scans taken along $\mathbf{Q} = (2, k ,0)$ and at $\mathbf{Q} = (1.5, 0, 0)$ at 300 and 140 K, to constant-$\mathbf{Q}$ cuts calculated from the MD S($\bm{Q},E$) as a way to validate the calculations. 

\subsection{Molecular dynamics simulations}\label{MD}

MD trajectories for deuterated MAPbI$_3$ are generated using identical simulations to those in ref. \cite{zhu2019mixed} with the mass of hydrogen set to that of deuterium. We used a $20\times20\times10$ super-cell that is based on the 12-atom pseudocubic unit cell with cell length $a\approx6.3 ~\text{\AA}$. The simulations were done in the \textsc{LAMMPS} package \cite{plimpton1995fast}. The inter-atomic potential is from ref. \cite{mattoni2015methylammonium}. Trajectories were integrated using a 0.5 fs time step throughout. For the first 20 ps, the system was thermalized in the NPT ensemble at 300 K and 0 Pa, then it was equilibrated for another 20 ps in the NVE ensemble. After equilibration, the equations of motion were integrated for 20 ns and trajectories were written to a file every 50 fs.

\subsection{S(Q,E) calculations}\label{calcsqe}

The dynamical structure factor, S($\bm{Q},E$), can be calculated from MD. Given a set of suitably accurate trajectories, $\bm{r}_i(t)$, S($\bm{Q},E$) can be evaluated directly as
\begin{equation}
    S(\bm{Q},E) = \Big \lvert \sum_i^N f_i(Q)\int \exp (i(\bm{Q}\cdot \bm{r}_i(t)-\frac{E}{\hbar}t))dt \Big \rvert^2.
    \label{eq:SQE_MD}
\end{equation}
See the supplementary information \cite{supp_info} and refs. therein \cite{dove1993introduction,brown2006intensity,hazemann2005high,squires1996introduction,harrelson2021computing,zushi2015effect,xiong2017native,van1954correlations} for a derivation of eq. \ref{eq:SQE_MD}. The theoretical $S(\bm{Q},E)$ calculations were done with a Python code developed by us \cite{sterling_pynamic}.  The calculated $S(\mathbf{Q},E)$ is weighted by X-ray atomic form factors or thermal neutron scattering lengths for direct comparison to experimental data \cite{hazemann2005high,brown2006intensity}. If the experiment uses neutrons, $f_i(Q)\equiv b_i$ is the $Q-$independent scattering length; if the experiment uses X-rays, $f_i(Q)$ is the $Q-$dependent atomic form factor where $Q$ is the magnitude of momentum transferred, $\bm{Q}$, from the incident beam of particles into the material. $E$ is the energy transferred into the material and the sum runs over all $N$ atoms. 

The reciprocal space resolution from MD is $\Delta Q_{H,K} \approx 0.05 ~ \text{\AA}^{-1}$ along the $H,~K$ directions and $\Delta Q_{L}\approx0.1~\text{\AA}^{-1}$ along the $L$ direction. We do not integrate the calculation over reciprocal space unless otherwise specified, so the pixels in theoretical S($\bm{Q}$,$E$) plots have this spacing. In order to ensure we are far from the transient thermalization regime in MD, we only sample the last nano-second of the simulation to calculate S($\bm{Q}$,$E$). The last nano-second is split into 20 blocks that are 50 ps long; S($\bm{Q}$,$E$) is then calculated from 10 non-consecutive blocks and averaged over them.  For 50 ps blocks with a 50 fs sampling interval, the energy resolution is about $0.08$ meV and the maximum resolvable energy is 41.4 meV.  

To compare to the experimental NDS data in Fig. \ref{fig:fig2}, theoretical S($\bm{Q}$,$E$) is integrated between $\pm0.5$ meV, which is comparable to the resolution in the experiment. To compare to the experimental XDS data in Fig. \ref{fig:fig2}, theoretical S($\bm{Q}$,$E$) is integrated between $\pm2$ meV.
To compare with INS data in Fig. \ref{fig:fig3}, both the experimental and theoretical S($\bm{Q}$,$E$) are integrated $\pm1$ meV. Anywhere that purely theoretical data is compared, the integration is $\pm1$ meV. For excitations [e.g. Fig. \ref{fig:fig3}(b)], theoretical $S(\mathbf{Q},E)$ is not integrated over energy; the MD energy resolution is set by the length of the trajectory sampled in the calculation.

The workflow and analysis outlined above was performed on MD simulations of protonated MAPbI$_3$ and no significant differences were observed. The deuterated results are used in the main text for direct comparison to the experimental measurements.

\subsection{Plotting}
The unit cells in Fig. \ref{fig:summary_fig}a, b were plotted in \textsc{vesta} \cite{momma2011vesta}. The volumetric plot in Fig. \ref{fig:summary_fig}e was made in \textsc{mayavi} \cite{ramachandran2011mayavi}. Most other plots were made using the matplotlib package for Python \cite{matplotlib}.

\section{Data and Software Availability} 
The experimental and computational datasets used in the analysis here will be made available upon reasonable request. INS data associated with this experiment will be made publicly available by the Science and Technology Facilities Council at DOI: 10.5286/ISIS.E.RB2010431 The code used to calculate $S(\bm{Q},E)$ is available for free online \cite{sterling_pynamic}.

\section{Author Contributions}
N.J.W., H.-G.S., and M.F.T. conceived the project. N.J.W., J.A.V., H.-G.S., F.Y., and M.J.K. performed the XDS and NDS experiments and data reduction and N.J.W., T.C.S., D.R., and M.F.T. performed the analysis. J.A.V., A.G.-P., I.C.S., and H.I.K. synthesized and mounted the single crystals. T.C.S. performed and analyzed the $S(\mathbf{Q},E)$ calculations from the MD datasets provided by B.A. and E.E. N.J.W., T.C.S., D.V., A.G-.P., I.C.S., P.M.G., and D.R. performed the INS experiments and contributed to the analysis. N.J.W. and T.C.S. prepared the figures and wrote the manuscript with contributions from all authors.

\backmatter

\bibliography{bibliography}
\end{document}


\title[Article Title]{Supplementary Material for ``The nature of dynamic local order in CH$_3$NH$_3$PbI$_3$ and CH$_3$NH$_3$PbBr$_3$"}

\author*[1,2]{\fnm{Nicholas} {J} \sur{Weadock}}\email{nicholas.weadock@colorado.edu}
\equalcont{These authors contributed equally to this work.}
\author[3]{\fnm{Tyler} {C} 
\sur{Sterling}}
\equalcont{These authors contributed equally to this work.}

\author[4,5]{\fnm{Julian} {A} \sur{Vigil}}

\author[5]{\fnm{Aryeh} \sur{Gold-Parker}}

\author[5]{\fnm{Ian} {C} \sur{Smith}}

\author[6]{\fnm{Ballal} \sur{Ahammed}}

\author[7]{\fnm{Matthew} {J} \sur{Krogstad}}

\author[8]{\fnm{Feng} \sur{Ye}}

\author[9,10]{\fnm{David} \sur{Voneshen}}

\author[{11}]{\fnm{Peter} {M} \sur{Gehring}}

\author[12]{\fnm{Hans-Georg} \sur{Steinr{\"u}ck}}

\author[6]{\fnm{Elif} \sur{Ertekin}}

\author[5,13]{\fnm{Hemamala} {I} \sur{Karunadasa}}

\author*[3]{\fnm{Dmitry} \sur{Reznik}}\email{dmitry.reznik@colorado.edu}

\author*[1,2,14]{\fnm{Michael} {F} \sur{Toney}}\email{michael.toney@colorado.edu}

\affil[1]{\orgdiv{Materials Science and Engineering}, \orgname{University of Colorado, Boulder}, \orgaddress{\city{Boulder}, \state{CO}, \postcode{80309}, \country{USA}}}

\affil[2]{\orgdiv{Department of Chemical and Biological Engineering}, \orgname{University of Colorado, Boulder}, \orgaddress{\city{Boulder}, \state{CO}, \postcode{80309}, \country{USA}}}

\affil[3]{\orgdiv{Department of Physics}, \orgname{University of Colorado, Boulder}, \orgaddress{\city{Boulder}, \state{CO}, \postcode{80309}, \country{USA}}}

\affil[4]{\orgdiv{Department of Chemical Engineering}, \orgname{Stanford University}, \orgaddress{\city{Stanford}, \state{CA}, \postcode{94305}, \country{USA}}}

\affil[5]{\orgdiv{Department of Chemistry}, \orgname{Stanford University}, \orgaddress{\city{Stanford}, \state{CA}, \postcode{94305}, \country{USA}}}

\affil[6]{\orgdiv{Department of Mechanical Science \& Engineering and Materials Resesarch Laboratory}, \orgname{University of Illinois at Urbana-Champaign}, \orgaddress{\city{Urbana}, \state{IL}, \postcode{61801}, \country{USA}}}

\affil[7]{\orgdiv{Advanced Photon Source}, \orgname{Argonne National Lab}, \orgaddress{\city{Lemont}, \state{IL}, \postcode{60439}, \country{USA}}}

\affil[8]{\orgdiv{Neutron Scattering Division}, \orgname{Oak Ridge National Laboratory}, \orgaddress{\city{Oak Ridge}, \state{TN}, \postcode{37830}, \country{USA}}}

\affil[9]{\orgdiv{ISIS Facility}, \orgname{Rutherford Appleton Laboratory}, \orgaddress{\city{Chilton}, \state{Didcot}, \postcode{Oxon OX11 0QX}, \country{United Kingdom}}}

\affil[10]{\orgdiv{Department of Physics}, \orgname{Royal Holloway University of London}, \orgaddress{\postcode{Egham TW20 0EX}, \country{United Kingdom}}}

\affil[11]{\orgdiv{NIST Center for Neutron Research}, \orgname{National Institute of Standards and Technology}, \orgaddress{\city{Gaithersburg}, \state{MD}, \postcode{20899}, \country{USA}}}

\affil[12]{\orgdiv{Department Chemie}, \orgname{Universt{\"a}t Paderborn}, \orgaddress{\city{Paderborn}, \country{Germany}}}

\affil[13]{\orgdiv{Stanford Institute for Materials and Energy Sciences}, \orgname{SLAC National Accelerator Laboratory}, \orgaddress{\city{Menlo Park}, \state{CA}, \postcode{94025}, \country{USA}}}

\affil[14]{\orgdiv{Renewable and Sustainable Energy Institute (RASEI)}, \orgname{University of Colorado, Boulder}, \orgaddress{\city{Boulder}, \state{CO}, \postcode{80303}, \country{USA}}}

\maketitle

\newpage

\section{Diffuse scattering of MAPbBr$_3$}

\begin{figure}
    \centering
    \includegraphics[width=0.95\textwidth]{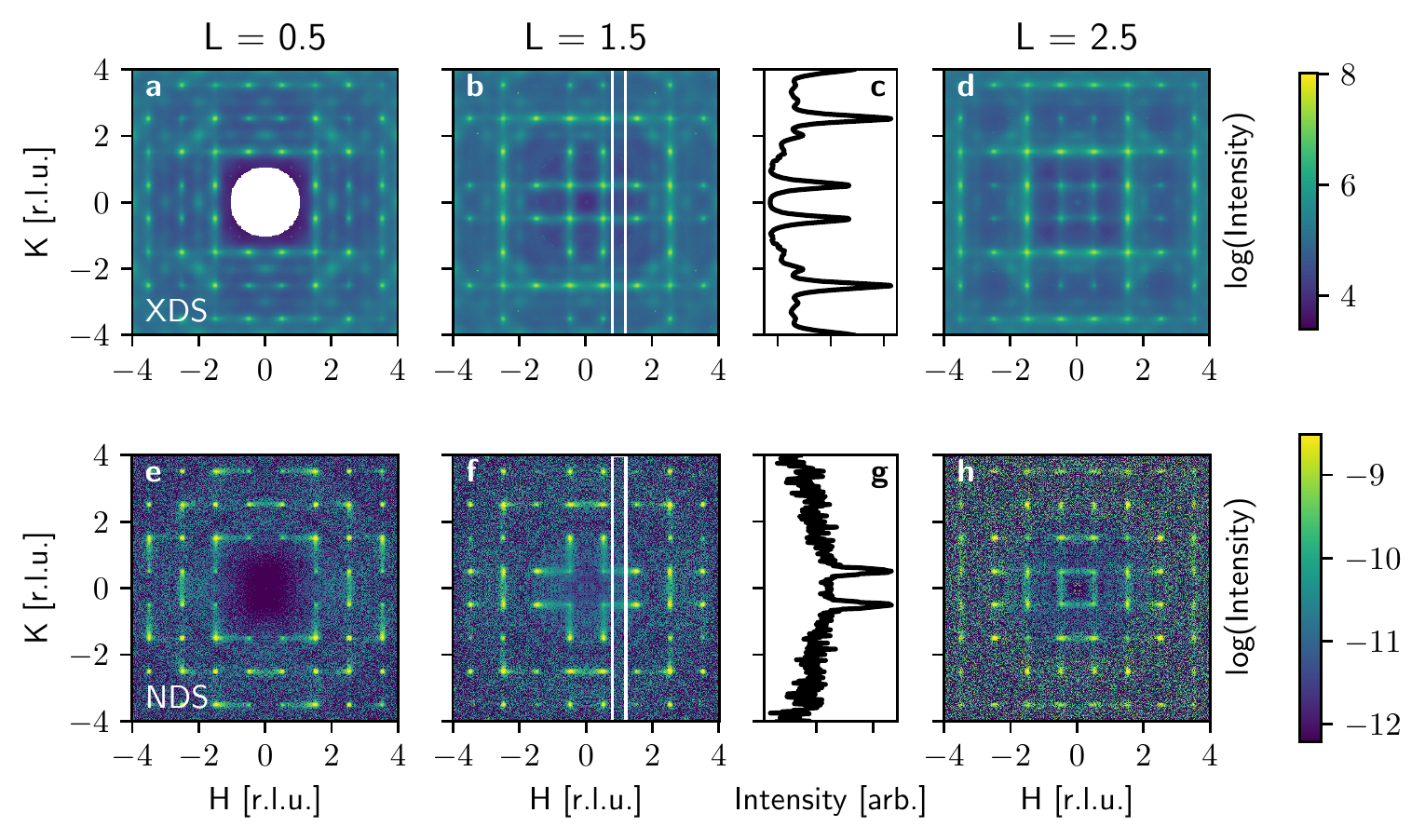}
    \caption{Experimental $S(\mathbf{Q})$ for protonated (a-d) and deuterated (e-h) MAPbBr$_3$ measured at 250 K (globally cubic phase) with X-ray and neutron diffuse scattering, respectively. The NDS data are limited to the elastic scattering channel of the CORELLI spectrometer (0.9 meV FWHM), whereas the XDS data are inherently energy integrated. Reciprocal space slices taken at L=0.5,1.5,2.5 contain only diffuse scattering intensity. Linecuts in c, g, are obtained by integrating within the white boxes in b, f, respectively, and illustrate the difference in diffuse scattering intensities in XDS and NDS. The intensity in a,b,d,e,f,h is plotted on a logarithmic scale.}
    \label{fig:S1}
\end{figure}

Representative slices of the reciprocal space volumes collected for MAPbBr$_3$ using X-ray (XDS) and neutron diffuse scattering (NDS) are plotted in Figure~S\ref{fig:S1}. Cuts in the H, K, L=0.5, 1.5, and 2.5 planes show diffuse scattering patterns nearly identical to MAPbI$_3$, namely extended (XDS) or modulated (NDS) rods of intensity along the Brillouin Zone (BZ) edge connecting the M and R-points. One-dimensional cuts in $\mathbf{Q}$ in Fig.~S\ref{fig:S1}c,g highlight the alternating intensity profile of NDS.

\section{S(Q,E) from molecular dynamics}\label{SI:SQW_MD}

We derive the equation used to calculate the dynamic structure factor, $S(\bm{Q}, E)$, from molecular dynamics.

The dynamic structure factor is defined as
\begin{equation}
    \begin{split}
    S(\bm{Q},\omega) & =\int \int \langle  \hat{\rho}(\bm{r}+\bm{r}',t+t') \hat{\rho}(\bm{r}',t') \rangle \exp(i(\bm{Q}\cdot \bm{r}-\omega t)) dt d\bm{r} \\
    & \equiv \int G(\bm{r},t) \exp (i(\bm{Q} \cdot \bm{r}-\omega t)) dt d\bm{r} \\
    & \equiv \int F(\bm{Q},t) \exp (-i \omega t) dt
    \end{split}
    \label{eq:dsf}
\end{equation}
Eq. \ref{eq:dsf} is the time- and space-Fourier transform of the density-density correlation function (also called the \emph{Van Hove} function \cite{van1954correlations}), $G(\bm{r},t)$. $\hat{\rho}(\bm{r},t)$ is the quantum mechanical density operator, by necessity defined using a quantum mechanical position operator $\hat{\bm{r}}_i(t)$.

Positions do not commute at different times $t$, $t'$ therefore evaluating Eq.~\ref{eq:dsf} is difficult. To simplify the notation, we do not write the explicit time dependence of $\hat{\bm{r}}$  except where it is needed. The usual method to evaluate Eq.~\ref{eq:dsf} for crystals is to expand the position operators, $\hat{\bm{r}}$, in terms of the phonon creation and annihilation operators \cite{squires1996introduction}. This works well when the harmonic approximation is sufficient, but at high-temperatures and particularly in molecular crystals where molecules rotate almost freely, this is not adequate. Instead, we approximate the positions as classical coordinates so that $\hat{\rho} \equiv \rho$ is classical and the classical positions $\bm{r}(t)$ can be determined using molecular dynamics simulations\footnote{Ab-initio molecular dynamics works as well because the atomic trajectories are classical.}. Importantly we have made no assumptions about the configuration of the material, so this computational technique is valid for liquids, disordered compounds, molecular crystals, etc. The main source of error of the classical approximation is that the scattering function $S(\bm{Q},\omega)$ does not satisfy the principle of detailed balance \cite{squires1996introduction,dove1993introduction,harrelson2021computing}. 

Still, we use the classical approximation for $\rho$ in all places where $S(\bm{Q},\omega)$ is calculated from molecular dynamics trajectories in the text. The classical approximation becomes valid at high-temperature where quantum effects on nuclear motion are negligible.

For classical (i.e. commuting) positions, Eq. \ref{eq:dsf} can be simplified. With
\begin{equation}
    \delta(\bm{r}-\bm{r}_i)=\int{\exp(-i\bm{Q}'\cdot(\bm{r}-\bm{r}_i))}\frac{d\bm{Q}'}{(2\pi)^3}
    \label{eq:delta_fn}
\end{equation}
we can write $\rho(\bm{r},t)$ as \cite{dove1993introduction}
\begin{equation}
    \rho(\bm{r},t)=\sum_i^N f_i(Q)\int{\exp(-i\bm{Q}'\cdot(\bm{r}-\bm{r}_i))} \frac{d\bm{Q}'}{(2\pi)^3}.
    \label{eq:rho_r}
\end{equation}
In Eq. \ref{eq:rho_r} $f_i(Q)\equiv b_i$ is the $\mathbf{Q}$-independent neutron scattering length if the experiment uses neutrons; if X-rays are used, then $f_i(Q)$ is the $\mathbf{Q}$-dependent atomic form factor which can be approximated by a sum of Gaussians
\begin{equation}
    f_i(Q)=\sum_j^4 p_{i,j} \exp \left(-q_{i,j} \left( \frac{Q}{4\pi}\right)^2\right)+s_i.
    \label{eq:fQ}
\end{equation}
The parameters $p_{i,j}$, $q_{i,j}$, and $s_i$ for X-rays and the scattering lengths $b_i$ for neutrons have already been determined and tabulated \cite{hazemann2005high,brown2006intensity}. The index $i$ runs over all atoms in the compound and $f_i(Q)$ is different for different atoms. $Q$ is the (magnitude of-) momentum transferred from the incident (mono-chromatic) beam into the sample. 

The classical expression for the Van Hove function, G($\bm{r},t$) in Eq. \ref{eq:dsf}, is
\begin{equation}
    G(\bm{r},t) = \langle \rho (\bm{r}+\bm{r}',t+t') \rho (\bm{r}',t') \rangle = \int \int \rho (\bm{r}+\bm{r}',t+t') \rho (\bm{r}',t') d\bm{r}'dt'.
    \label{eq:Grt_1}
\end{equation}
Inserting Eq. \ref{eq:rho_r} into Eq. \ref{eq:Grt_1} and carrying out the integrals, we find
\begin{equation}
    G(\bm{r},t)=\sum_i^N \sum_j^N f_i(Q) f_j(Q) \int  \delta(\bm{r}-(\bm{r}_i(t+t')-\bm{r}_j(t'))) dt' .
    \label{eq:Grt}
\end{equation}
Similarly, inserting Eq. \ref{eq:Grt} into Eq.~\ref{eq:dsf}, we find:
\begin{equation}
    F(\bm{Q},t) = \sum_i^N \sum_j^N f_i(Q) f_j(Q) \int \exp (i\bm{Q}\cdot(\bm{r}_i(t+t')-\bm{r}_j(t'))) dt'.
    \label{eq:FQt}
\end{equation}
Next, we can rewrite $\exp(i\bm{Q}\cdot\bm{r}(t))$ as
\begin{equation}
    \exp(i\bm{Q}\cdot\bm{r}(t))=\int \exp(i\bm{Q}\cdot\bm{r}(\tau)) \delta(\tau-t)d\tau.
    \label{eq:exp_tau}
\end{equation}
Combining equations \ref{eq:FQt} and \ref{eq:exp_tau} with Eq. \ref{eq:dsf}, we can do all the integrals over exponentials:
\begin{equation}
    \begin{split}
    S(\bm{Q},\omega) = \sum_i^N \sum_j^N f_i(Q) f_j(Q) \int \int & \exp (i(\bm{Q}\cdot \bm{r}_i(\tau)-\omega\tau))  \times  \\
    & \exp (-i(\bm{Q}\cdot \bm{r}_j(\tau')-\omega\tau')) d\tau d\tau' .
    \end{split}
    \label{eq:SQw}
\end{equation}
Finally, with $E\equiv\hbar\omega$ and $\tau\equiv t$, we can rewrite this as
\begin{equation}
    S(\bm{Q},E) = \Big \lvert \sum_i^N f_i(Q)\int \exp (i(\bm{Q}\cdot \bm{r}_i(t)-\frac{E}{\hbar}t))dt \Big \rvert^2.
    \label{eq:A_SQE_MD}
\end{equation}
Equation \ref{eq:A_SQE_MD} can be evaluated from molecular dynamics trajectories, $\bm{r}_i(t)$. An expression similar to Eq. \ref{eq:SQw} has been used in the past \cite{zushi2015effect,xiong2017native}. 

\section{How well do MD and experiment agree?}
\begin{figure}
\centering
\includegraphics[width=0.8\linewidth]{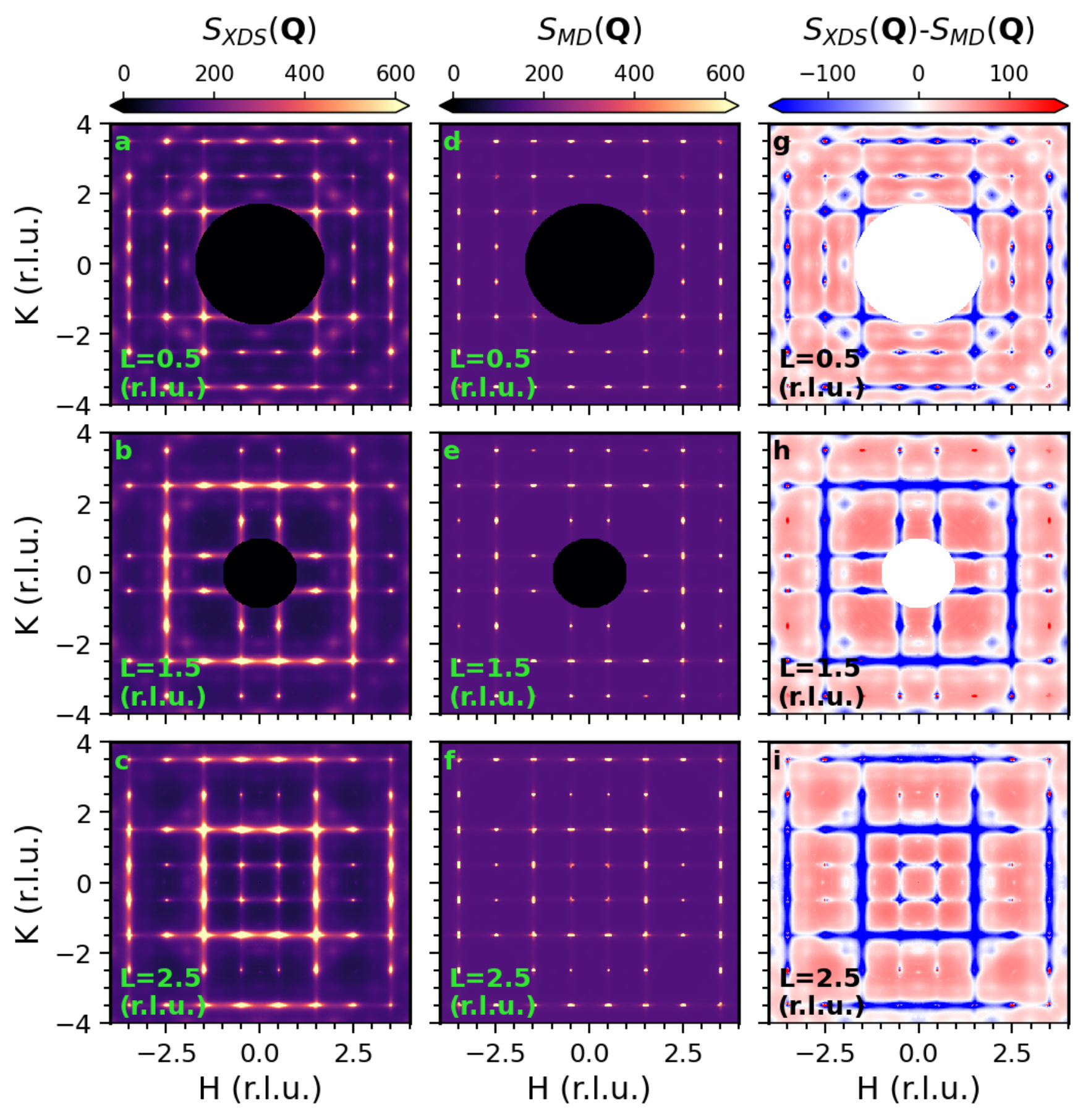}
    \caption{Difference between XDS for MAPbI$_3$ at 345 K and ``best-fit" MD data at 300 K. The MD data in d-f have a constant offset of $151.25$ and scale of $40.22$, determined by minimizing the RMS error with these as parameters. The RMS error is $91.78$ ($3.9\%$ of the maximum, 2164.92). The error, $S_{\mathrm{XDS}}(\bm{Q})-S_{\mathrm{MD}}(\bm{Q})$, for each $L$ are in g-i.}
\label{SI_fig:error}
\end{figure}

\begin{figure}
\centering
\includegraphics[width=1\linewidth]{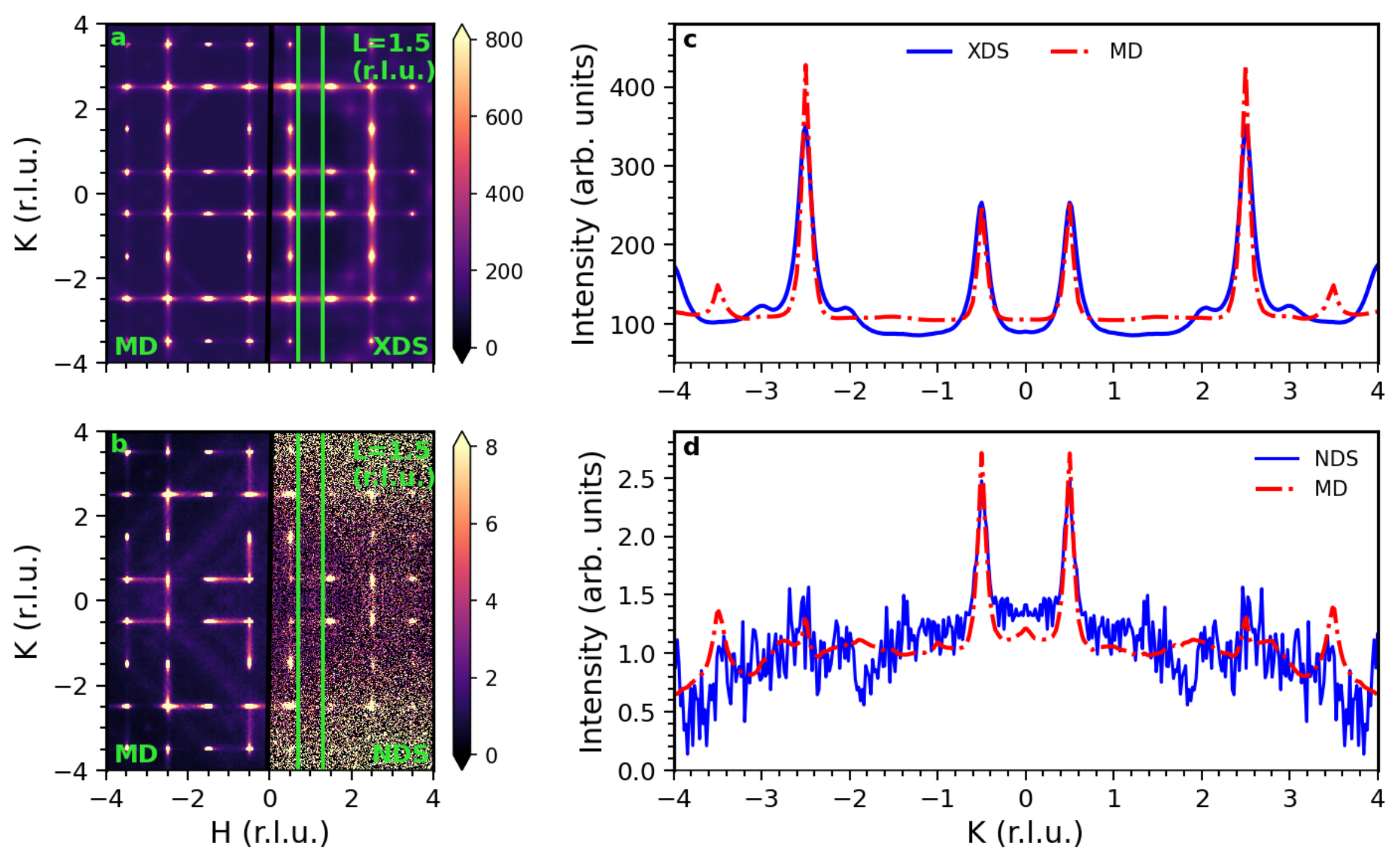}
    \caption{Linecut across the diffuse rods highlighting the agreement between MD and experiment. a Diffuse scattering from MAPbI$_3$ in the $L=1.5$ plane from MD (left) and XDS at 345 K (right). c A linecut through the intensity in a. Data in c are integrated over the region between the vertical lines in a [$0.7<H<1.3$, $K$, $1.4<L<1.6$ r.l.u.]. The height and scale of the MD data in c were fit to the XDS data with an RMS error of 6.8$\%$. b Diffuse scattering in the $L=1.5$ plane from MD (left) and MAPbI$_3$ NDS at 345 K (right). d A linecut through the same region as in c comparing calculated and MD NDS data. The height and scale of the MD data in d were fit to the NDS data with an RMS error of 10.1$\%$. The MD data in a and b are scaled using the results of the fits in c and d. In c, the peaks at integer values of K in the XDS data are likely phonons.}
\label{SI_fig:linecut_error}
\end{figure}

The relative scale of the theoretical and experimental $S(\bm{Q},E)$ is arbitrary since there are no incident X-ray or neutron fluxes in the MD simulation and this quantity was not explicitly measured in the experiments. Moreover, there are background contributions from the experimental setup that are not present in MD. In the comparisons between the calculated and experimental $S(\bm{Q},E)$ in the main text, the calculated data are multiplied by an arbitrary scaling factor to account for these differences and to allow a qualitative comparison. To obtain a more quantitative measure of the accuracy of our MD results, we calculate the root-mean-squared (RMS) error between experimental and computed $S(\bm{Q})$, adjusted for a constant background and scaled for incident flux. First, we consider the agreement in two dimensions for the L = 0.5, 1.5, and 2.5 scattering planes. The background offset and scale are varied to minimize the RMS error. Data in a sphere around $\bm{Q}=(0,0,0)$ were removed from the MD data because there are no data here in the XDS experiment. The results of ``fitting" the MD data to experiment in the L = 0.5, 1.5, 2.5 r.l.u. planes are shown in Fig.~S\ref{SI_fig:error}. The scale and offset are summarized in the caption.

In Fig.~S\ref{SI_fig:error} we calculate the RMS error as $91.78$ ($3.9\%$ of the maximum in XDS data, 2164.92). A particular source of error is the ``empty" space between diffuse rods; the experimental data are not constant here, but we approximate the background as constant in the comparison. Comparing the data near the zone-boundaries shows that the $R$-points in the MD data are more intense than the diffuse-rods: the $R$-points are very bright in the MD figures (linear scale), while the rods are almost invisible. This is apparent in Fig.~S\ref{SI_fig:error}g-i, where the intensity at the $R$-points is too large, and the intensity near the diffuse rods is too low. This is likely an artifact of the difference in simulation temperature (300 K) and measurement temperature (345 K and above), as the R-point intensity increases with decreasing temperature \cite{cominLatticeDynamicsNature2016, beecherDirectObservationDynamic2016,  weadockTestDynamicDomainCritical2020}.

We also compare MD data and experiment using linecuts across diffuse rods in Fig. S\ref{SI_fig:linecut_error}. In this case, the fitting results in RMS error of 6.8$\%$ and 10.1$\%$ for XDS and NDS, respectively. The MD is scaled by the parameters obtained in the linecut fits and representative scattering in the L = 1.5 plane is presented in Fig. S\ref{SI_fig:linecut_error} a and b. Qualitatively, the MD and XDS, NDS have better visual agreement than in Fig. S\ref{SI_fig:error} in which whole planes are considered for the fit. In particular, we find that the diffuse rods look nearly identical between MD and experiment in Fig. S\ref{SI_fig:error}a. Since we are mainly interested in the diffuse rods, in the main text we vary the scale and background of the calculated $S(\bm{Q})$ to highlight these features. We note that scaling by the Bragg intensity is not possible as the Bragg intensities measured in XDS can saturate the Pilatus detector pixels and therefore does not provide an accurate measure of the intensity.

\begin{figure}
    \centering
    \includegraphics[width = \textwidth]{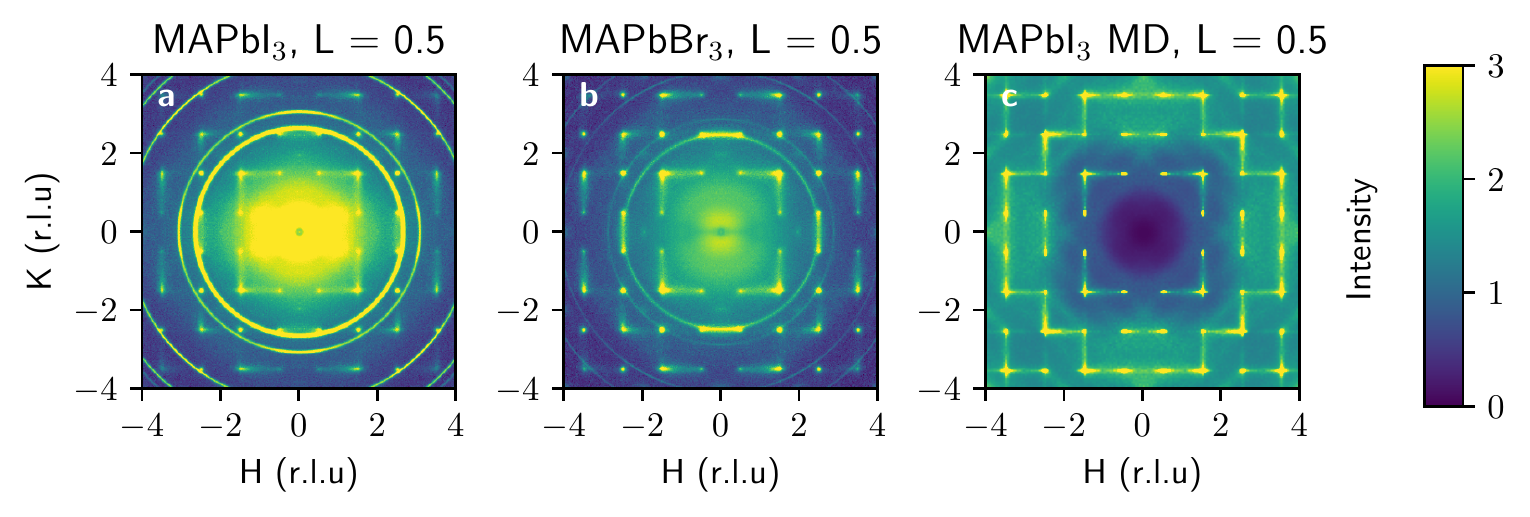}
    \caption{Experimental neutron diffuse scattering in the L = 0.5 plane for a, MAPbI$_3$ at 345 K; b, MAPbBr$_3$ at 250 K; and c, MD calculated NDS for MAPbI$_3$. In a,b, we note that the diffuse intensity along the rods decays away from one of the two R-point endpoints. The Debye-Scherrer powder rings from the aluminum are subtracted in Fig. 2 in the main text and Fig. S\ref{fig:S1} for clarity but do not alter the diffuse scattering intensity.}
   
    \label{fig:NDScompare}
\end{figure}

We qualitatively compare the experimental and calculated NDS in Fig.~S\ref{fig:NDScompare}. A quantitative comparison is difficult due to the significant Debye-Scherrer rings from the aluminum sample holder used in NDS. The alternating nature of the intensity along the diffuse rods is well reproduced for smaller H, K, values in the L = 0.5 plane, however there is reduced intensity in the experimental NDS for rods originating from the $\mathbf{Q} = (3.5,3.5,0.5)$ and $\mathbf{Q} = (1.5,3.5,0.5)$ families of R-points. This is apparent for both MAPbI$_3$ and MAPbBr$_3$, and indicates that the intermolecular correlations are not as strong as predicted by the MD. Our simulations utilize a classical, electrostatic potential which has been demonstrated to accurately, but not perfectly, model the experimental molecular relaxation times of MA$^+$ \cite{mattoni2015methylammonium}. The simulation also does not capture defects or thermally excited carriers which could screen electrostatic interactions between MA$^+$. 

\section{Intensity variation along the rods}

\begin{figure}
\centering
\includegraphics[width=1\linewidth]{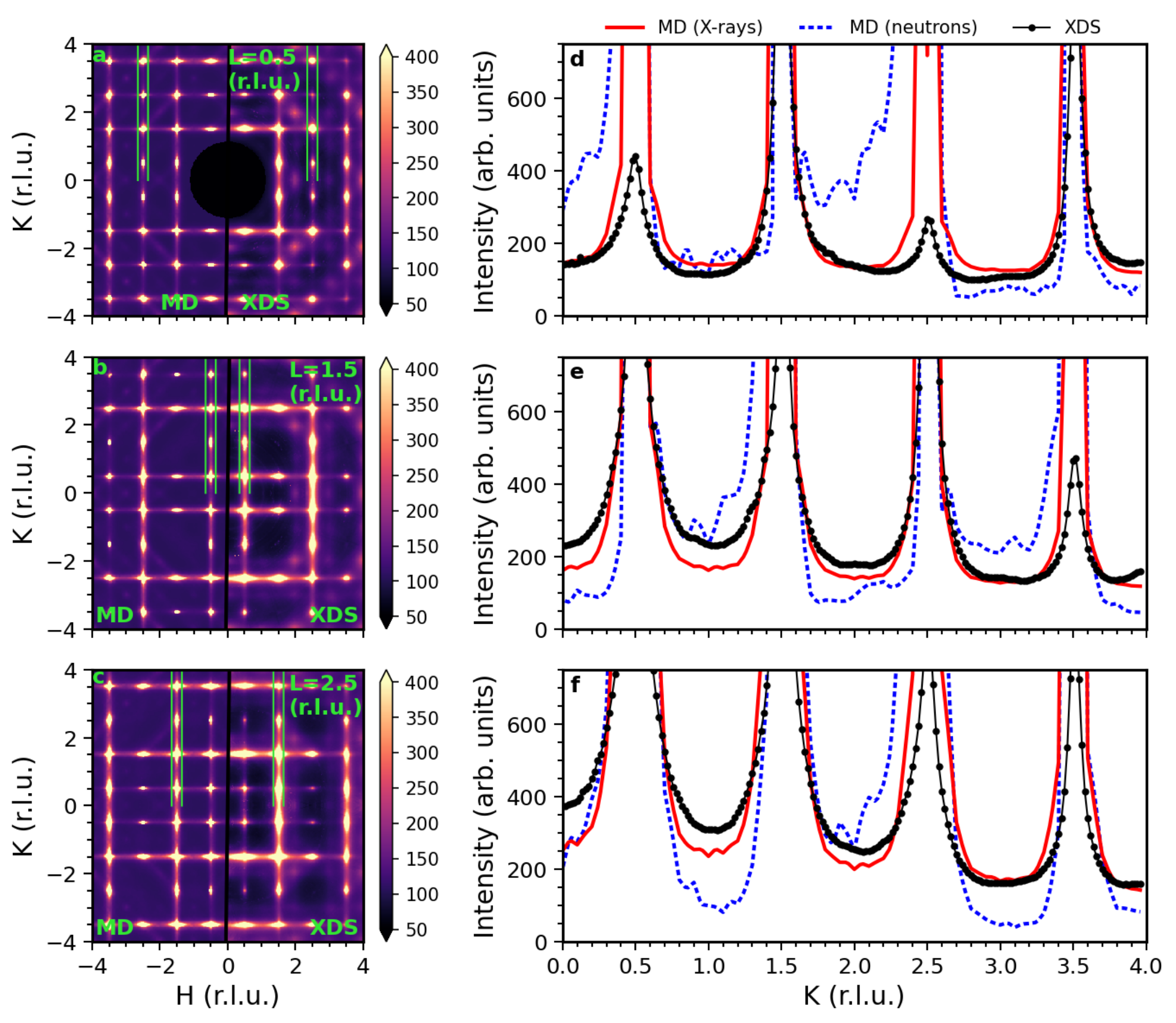}
    \caption{Comparison between the diffuse intensity along the rods in in MAPbI$_3$ from MD and XDS at 345 K. The MD data (with X-ray form-factors) are on the left and XDS data on the right in a-c. Linecuts integrated between the green lines in a-c are plotted in d-f; MD (X-ray) data are red lines, MD (with neutron scattering lengths) are blue dotted lines, and the XDS data are black dots. MD (neutron) data are not plotted in a-c but provided for comparison. See Fig.~\ref{SI_fig:diffuse_MD} for the corresponding data. The shift and scale applied to the MD data is the same as determined in Fig. S\ref{SI_fig:linecut_error}. The XDS data in d-f are integrated $\pm0.5$ r.l.u. while the MD are not integrated over momentum.}
\label{SI_fig:along_rods_xrays}
\end{figure}

We note in the main text that the XDS and NDS diffuse rods have qualitatively different intensity profiles: the magnitude of XDS intensity is relatively constant along several BZs, whereas for NDS the intensity is reduced for every other BZ as observed in Figs.~S\ref{fig:S1} and S\ref{fig:NDScompare}. Furthermore, along individual BZ edges the NDS diffuse intensity decays from the R-point (Fig.~S\ref{fig:NDScompare}). By contrast, the XDS intensity is constant except for the R-point contribution. Fig.~S\ref{SI_fig:along_rods_xrays} plots linecuts along the diffuse rods for the 354 K XDS data along $\bf{Q}$ = (2.5, k, 0.5), (0.5, k, 1.5) and (1.5, k, 2.5) in a,b,c, respectively. Linecuts along the rods extracted from the MD X-ray and neutron $S(\bm{Q})$ are also plotted for comparison. We see the experimental and MD XDS intensity is constant between the R-points, whereas the NDS intensity falls off asymmetrically between the R-points. Since XDS shows no variation along the rod and NDS does, the variation in NDS likely arises from scattering from the MA$^+$ molecules which are nearly invisible to X-rays. We verify this below. 

The sublattice contributions to the diffuse scattering intensity are analyzed in the following way. Denoting $\rho(\bm{Q},E)=\sum_{i} f_i(Q)\int \exp (i(\bm{Q}\cdot \bm{r}_i(t)-tE/\hbar))dt$ with $i$ labeling the atom and the sum running over all atoms in a particular group, we separate the intensity into terms corresponding to the inorganic PbI$_6$ (cage) and organic (MA$^+$) sublattices:
\begin{equation}
\begin{split}
    S(\bm{Q},E) & = \lvert \rho_{\text{MA}}(\bm{Q},E) + \rho_{\text{cage}}(\bm{Q},E) \rvert^2 \\
    & = \lvert \rho_{\text{MA}}(\bm{Q},E) \rvert^2 + \lvert \rho_{\text{cage}}(\bm{Q},E) \rvert^2 \\
    &+ \rho_{\text{MA}}^*(\bm{Q},E)\rho_{\text{cage}}(\bm{Q},E) +\rho_{\text{cage}}^*(\bm{Q},E)\rho_{\text{MA}}(\bm{Q},E)
   \end{split}
\end{equation}
where $\rho_{\text{MA}}(\bm{Q},E)$ and $\rho_{\text{cage}}(\bm{Q},E)$ are only summed over the atoms in the groups denoted by the subscript: MA$^+$ (methylammonium) are C, D (deuterium), and N atoms while cage atoms are Pb and I. We call $\lvert \rho_{\text{MA}}(\bm{Q},E=0) \rvert^2\equiv S_\text{MA}(\bm{Q})$ and similarly for the cage atoms. The interference term is calculated from $S_\text{int.}(\bm{Q})=S(\bm{Q})-(S_\text{MA}(\bm{Q})+S_\text{cage}(\bm{Q}))$.

\begin{figure}
\centering
\includegraphics[width=1\linewidth]{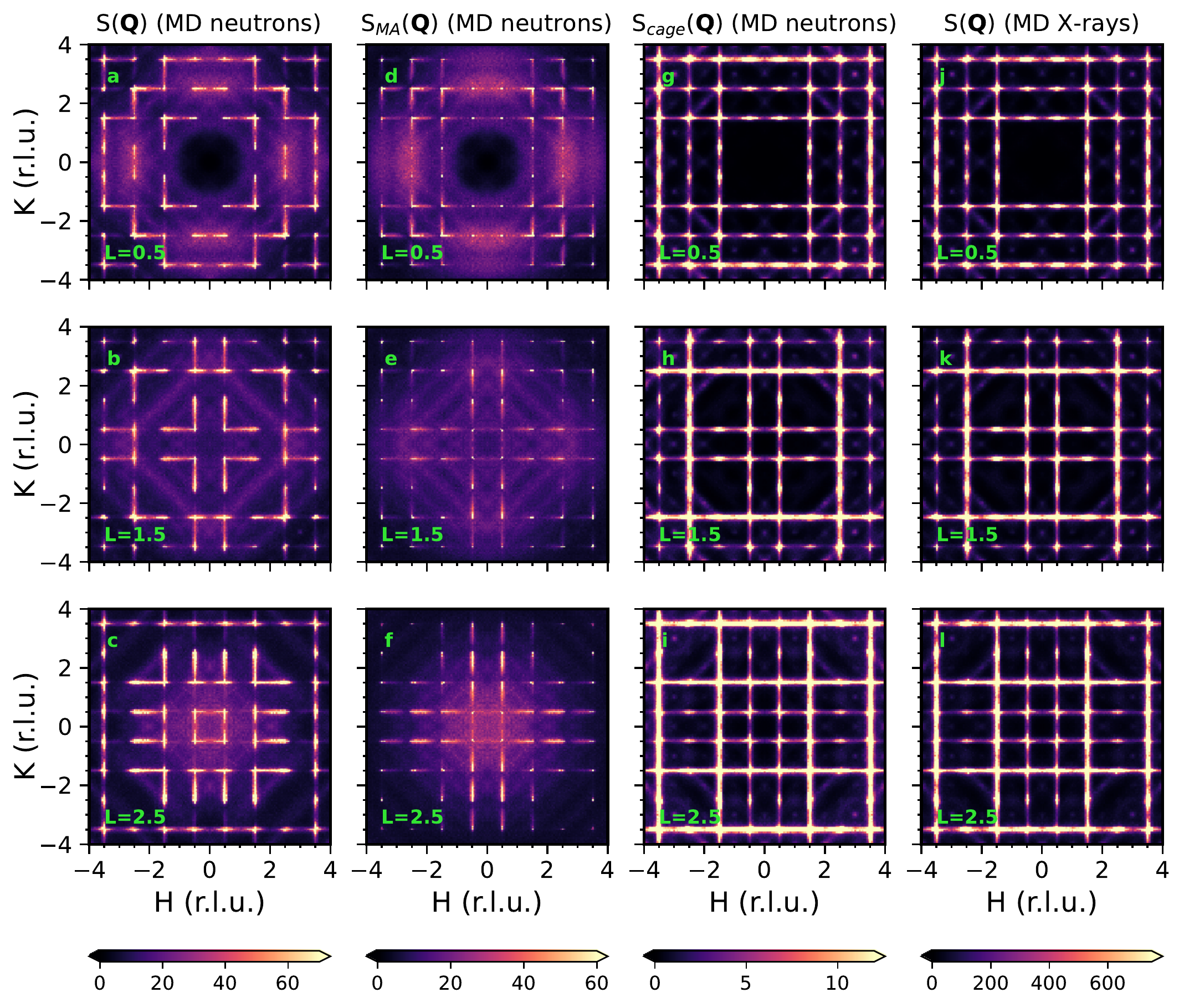}
    \caption{Different sublattice contributions to the total diffuse scattering intensity calculated from MD. The total NDS intensity is in a-c and the total XDS intensity is in j-l. Scattering from only the MA and cage sublattices is in d-f and g-i, respectively.}
\label{SI_fig:diffuse_MD}
\end{figure}

\begin{figure}
\centering
\includegraphics[width=1\linewidth]{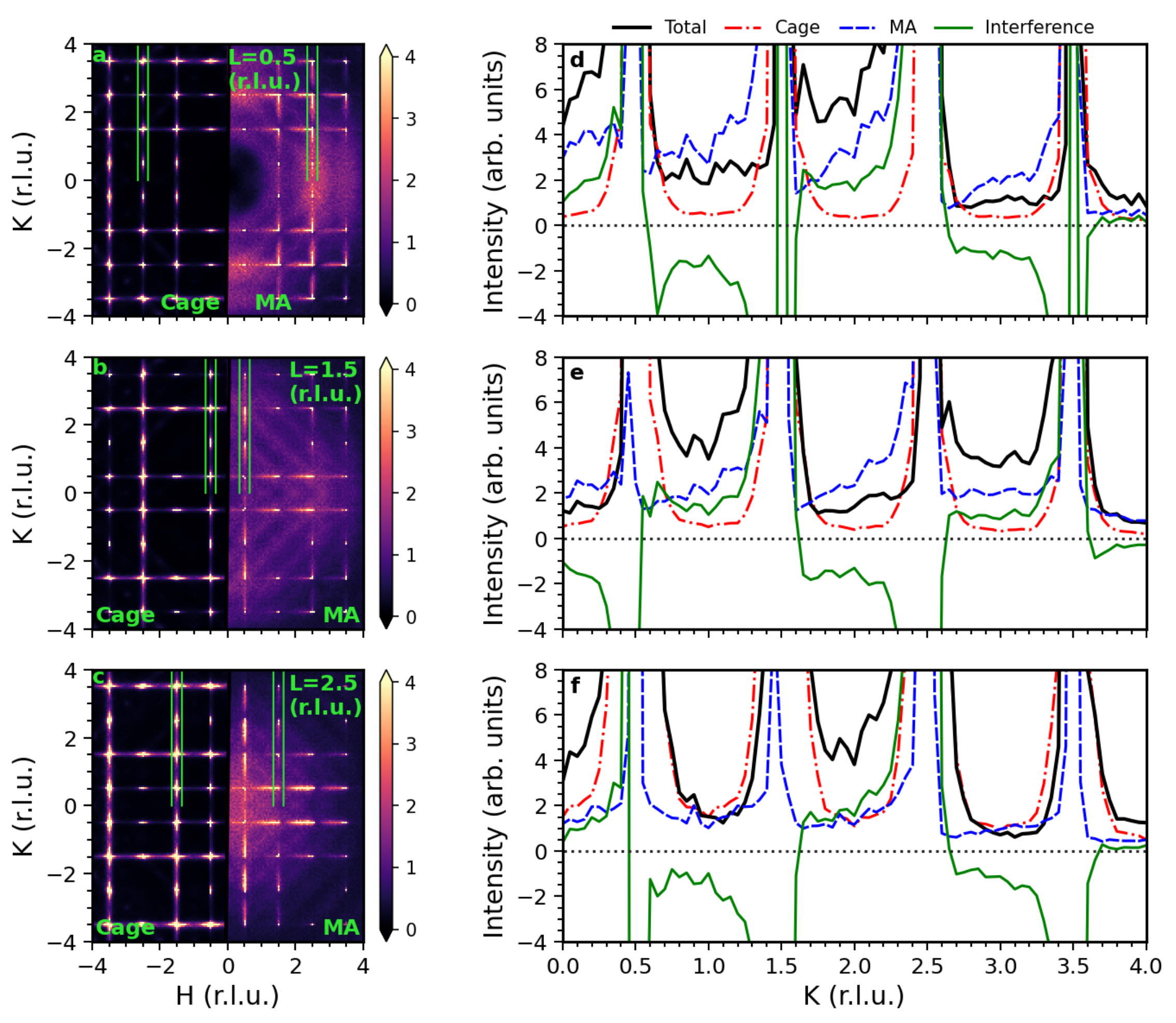}
    \caption{Different sublattice contributions to the intensity along the diffuse rods in MAPbI$_3$ from MD. The cage and MA$^+$ sublattice contributions to the theoretical neutron scattering intensity, $S_\text{cage}(\bm{Q})$ and $S_\text{MA}(\bm{Q})$, are on the left and right respectively in a-c. In d-f, the cage and MA$^+$ contributions are labeled in the legend above. Also shown are the total intensity and interference components from MD. The MD data are integrated $\pm0.5$ meV.} 

\label{SI_fig:along_rods_neutrons}
\end{figure}

The different contributions to $S(\bm{Q})$ are plotted in Fig.~S\ref{SI_fig:diffuse_MD}. We find that $S_\text{cage}(\bm{Q})$ is nearly indistinguishable from the calculated X-ray intensity (Fig. S\ref{SI_fig:diffuse_MD} g-i).
i.e. the XDS measurement only probes the correlations of the PbI$_6$ octahedra. We also extract linecuts from each $S(\bm{Q})$ contribution along $Q$ = (2.5, k, 0.5), (0.5, k, 1.5) and (1.5, k, 2.5) and plot them in Fig.~S\ref{SI_fig:along_rods_neutrons}. The linecuts corresponding to $S_\text{cage}(\bm{Q}))$ are in good agreement with the experimental and MD XDS in Fig.~S\ref{SI_fig:along_rods_xrays}. Moreover, because the XDS and cage-only intensities show these rods of constant intensity (Figs.~S\ref{SI_fig:along_rods_xrays},S\ref{SI_fig:along_rods_neutrons}), it is clear that the PbI$_6$ octahedra structural correlations have a two-dimensional character. 

The plots of $S_\text{MA}(\bm{Q})$ in Figs. S\ref{SI_fig:diffuse_MD} and S\ref{SI_fig:along_rods_neutrons} show two components of MA$^+$ diffuse scattering. First is the broad, spherical component corresponding to uncorrelated disorder arising from such molecular reorientations as rotations of methyl or amine groups about the C-N axis of MA$^+$. This component manifests as the nearly isotropic, circular intensity in Fig. S\ref{SI_fig:diffuse_MD}d-f. Second are the strong, distinct diffuse rods with varying intensity along the rod direction. The rods orient parallel to one of the H,K,L directions and indicate the presence of intermolecular structural correlations similar to the pancakes of tilted PbI$_6$ octahedra. The varying intensity, however, implies that the intermolecular correlations extend beyond one unit cell in the out-of-plane direction. Due to the nonzero $S_\text{int.}(\bm{Q})$ term, there is no clear way to isolate intermolecular MA$^+$ correlation lengths from the experimental or calculated diffuse scattering. Instead, we evaluate it explicitly below from the MD trajectories.

\begin{figure}
\centering
\includegraphics[width=1\linewidth]{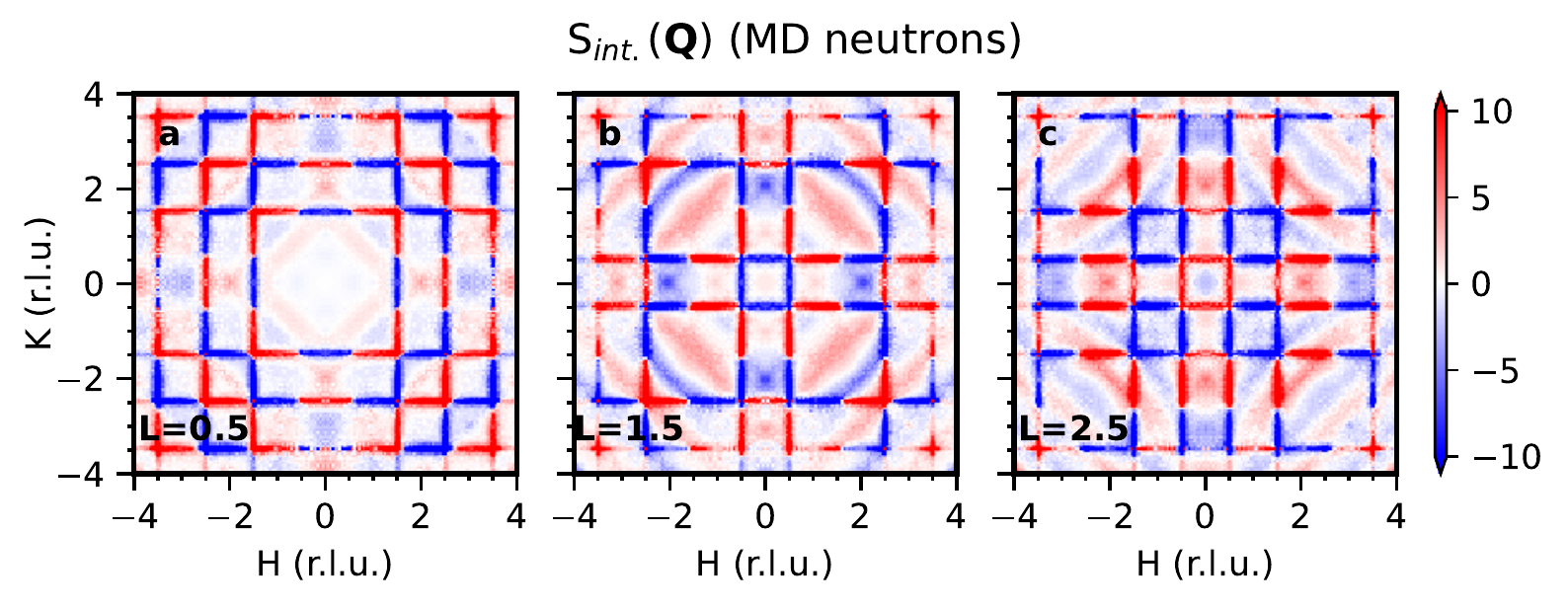}
    \caption{The interference part of S($\bm{Q}$) calculated from MD. The data in a are in the $L=0.5$ r.l.u plane, in b are in the $L=1.5$ r.l.u plane, and in c are in the $L=2.5$ r.l.u plane.}
\label{SI_fig:cross_term}
\end{figure}

The interference term $S_\text{int.}(\bm{Q})$ is plotted in Figs.~S\ref{SI_fig:along_rods_neutrons}, S\ref{SI_fig:cross_term} and indicates that the separate sublattice correlations are connected to one another. If they were entirely uncorrelated, $S_\text{int.}(\bm{Q})= 0$ everywhere and the total $S(\bm{Q}) = S_\text{cage}(\bm{Q}) + S_\text{MA}(\bm{Q})$ . Instead, there are regions of constructive and destructive interference, with the constructive component resembling the total NDS intensity (a-c in Fig. S\ref{SI_fig:diffuse_MD}). We propose that the structural correlations between the two sublattices result from MA$^+$ molecules reorienting into the lowest energy configuration in the cubocathedral region between PbI$_6$ octahedra, as discussed in the Main Text.

\section{Correlation lengths}

Correlation lengths $\xi$, corresponding to the widths of the two-dimensional structural correlations, are evaluated as stated in the Main Text and reported in Table~S\ref{tab:cl} below. For XDS experiments the $\xi$ correspond to PbX$_6$ correlations only, because the X-ray scattering intensity from light elements like C, N, D is significantly weaker.

\begin{table}[h]
\begin{center}

\caption{Correlation lengths obtained from $\mathbf{Q}$ linewidths of the diffuse rods and R-points. We report average values and standard deviation determined from several diffuse rods or R-points. For all R-point linewidths fitted here, contributions from the diffuse rod intensity have been subtracted prior to fitting. All reported lengths are in $\textrm{\AA}$, and correspond to the radius of the structural correlation.}\label{tab1}%
\begin{tabular}{|c|c|c|c|c|c|c|c|}
\toprule
 & \multicolumn{2}{c|}{MAPbBr$_3$} & \multicolumn{5}{c|}{MAPbI$_3$} \\
\midrule
 & \multicolumn{2}{c|}{250 K} & 330 K & 335 K & \multicolumn{2}{c|}{345 K} & 360 K\\
 & XDS\footnotemark[1] & NDS & XDS & NDS & XDS & NDS & XDS\\
\midrule
Diffuse rods& $10(2)$ & $10(2)$ & $13(4)$ & $15(4)$ & $12(3)$ & $15(8)$ & $11(3)$\\
R-points& $19(2)$ & $17(7)$ & $23(4)$ & $25(7)$ & $19(4)$ & $21(6)$ & $17(2)$\\
\botrule
\end{tabular}
\label{tab:cl}
\footnotetext[1]{Experiments performed on protonated MAPbBr$_3$.}
\end{center}
\end{table}

In addition to determining the PbI$_6$ correlation lengths from the diffuse scattering data, we evaluate the PbI$_6$ and MA$^+$ $\xi$ directly from the MD trajectories. To do this, we define a general correlation function between two dynamical variables, $\alpha(\bm{r},t)$ and $\beta(\bm{r},t)$ as \cite{allen2017computer}
\begin{equation}
\begin{gathered}
    G_{\alpha \beta}(\bm{r},t) = \langle \alpha(\bm{r},t) \beta(0,0) \rangle = \int \alpha(\bm{r}',t') \beta(\bm{r}+\bm{r}',t+t') d\bm{r}' dt'. 
    \label{eq:corr_func}
\end{gathered}
\end{equation}
The space-integral at equal time extends over the entire volume of the simulation box and the time integral over a long enough period to sample the correlations adequately. For a system with long-range order, $G_{\alpha \beta}(\bm{r},t)$ will be finite at infinite distance, whereas a system with short-range order exhibits a peak which decays towards zero at large $\bm{r}$ and $t$.

\begin{figure}
\centering
\includegraphics[width=0.4\linewidth]{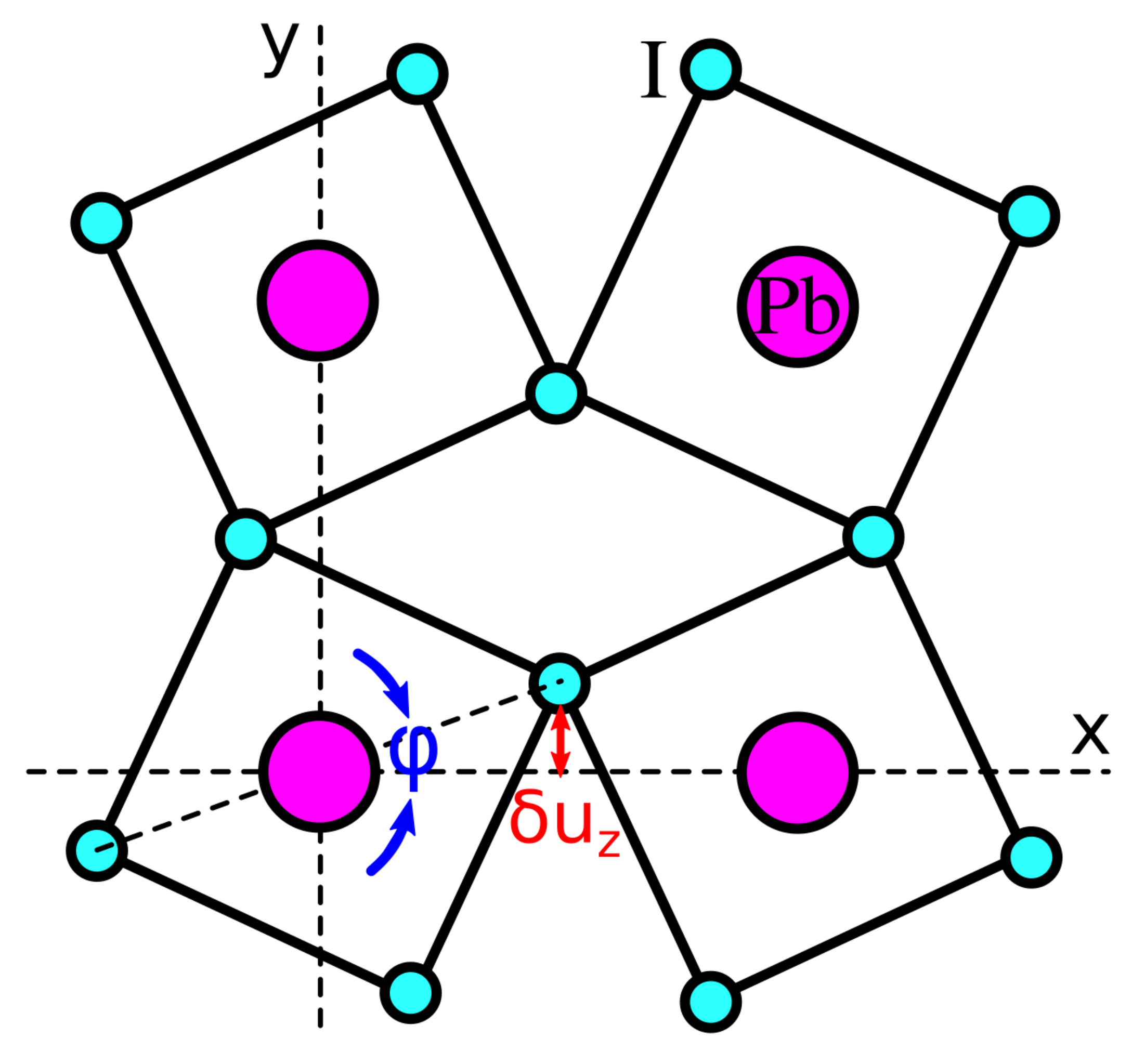}
    \caption{Diagram depicting the dynamical variable $\delta u_z(\bm{r},t)$ for explicit calculation of the PbI$_6$ correlations. $\delta u_z$ are the displacements of the iodine atoms due to rotation of the octahedra about the $z$-axis by angles $\phi$.} 
\label{SI_fig:iodine_displacement}
\end{figure}

For the PbI$_6$ correlations, we use the ``in-plane" iodine atom displacements, $\delta u_z(\bm{r},t)$, as dynamical variables (see Fig. S\ref{SI_fig:iodine_displacement}). The in-plane displacement is defined as that of the iodine atom perpendicular to the Pb-I bond arising from rotations of the octahedra about the axis perpendicular to the plane. We limit our analysis to displacements in the $x-y$ plane; i.e. rotations about the $z-$axis, but note that these correlations are present within the $y-z$ and $z-x$ planes as well, as discussed in the Main Text. We define $\delta u_z(\bm{r},t) = \sum_i \delta(\bm{r}-\bm{r}_i) \sin(\phi_i(t))$ with the bond-length set to unity and the octahedra fixed at the center of the unitcell. Thus, the PbI$_6$ correlation function only incorporates azimuthal rotations of the octahedra. 

In the case of the MA$^+$ cations, the variables $\alpha(\bm{r},t)$ and $\beta(\bm{r},t)$ are the Cartesian components of the MA$^+$ molecular orientation: $\bm{n}(\bm{r},t) =  n_x(\bm{r},t) \bm{\hat{x}} + n_y(\bm{r},t) \bm{\hat{y}} + n_z(\bm{r},t) \bm{\hat{z}}$. The vectors $\bm{n}_i(t)$ are parallel to the relative position vector between the C and N atoms belonging to the same MA$^+$ cation. The orientation field is $\bm{n}(\bm{r},t)\equiv \sum_i \bm{n}_i(t) \delta(\bm{r}-\bm{r}_i)$ with $\bm{r}_i$ the center-of-mass coordinate of the molecule. The sum runs over all MA$^+$ cations in the simulation cell (using periodic boundary conditions). For simplicity, we also assume $\bm{n}_i(t)$ are dimensionless unit vectors and that their centers-of-mass are fixed at the unit cell center. This isolates the MA$^+$ correlation function to measuring the MA$^+$ orientations only. 

Eq. \ref{eq:corr_func} can be evaluated directly, but the computational cost scales poorly with system size and time interval. It is more efficient to evaluate in reciprocal space where, by the convolution theorem, convolution (e.g. eq. \ref{eq:corr_func}) becomes simple multiplication. The reciprocal space analogue of eq. \ref{eq:corr_func} is 
\begin{equation}
\begin{gathered}
    G_{\alpha \beta} (\bm{q},\omega) = \alpha^*(\bm{q},\omega)  \beta(\bm{q},\omega) .
    \label{eq:corr_func_Qw}
\end{gathered}
\end{equation}
$\alpha(\bm{q},\omega)$ is the space and time Fourier transform of the $\alpha(\bm{r},t)$ and similarly for $\beta(\bm{q},\omega)$. $^*$ denotes complex conjugation. The real space version is recovered by inverse Fourier transforming (assuming the volume and time interval are infinite for notational convenience):
\begin{equation}
\begin{gathered}
    G_{\alpha \beta}(\bm{r},t) = \int G_{\alpha \beta} (\bm{q},\omega)  \exp(-i\bm{q}\cdot\bm{r}+i\omega t) \frac{d\omega }{2\pi} \frac{d\bm{q}}{(2\pi)^3} .
    \label{eq:corr_func_fft}
\end{gathered}
\end{equation}
Eq. \ref{eq:corr_func_fft} is efficiently evaluated using numerical fast Fourier transforms. 

\begin{figure}
    \centering
    \includegraphics[width = \columnwidth]{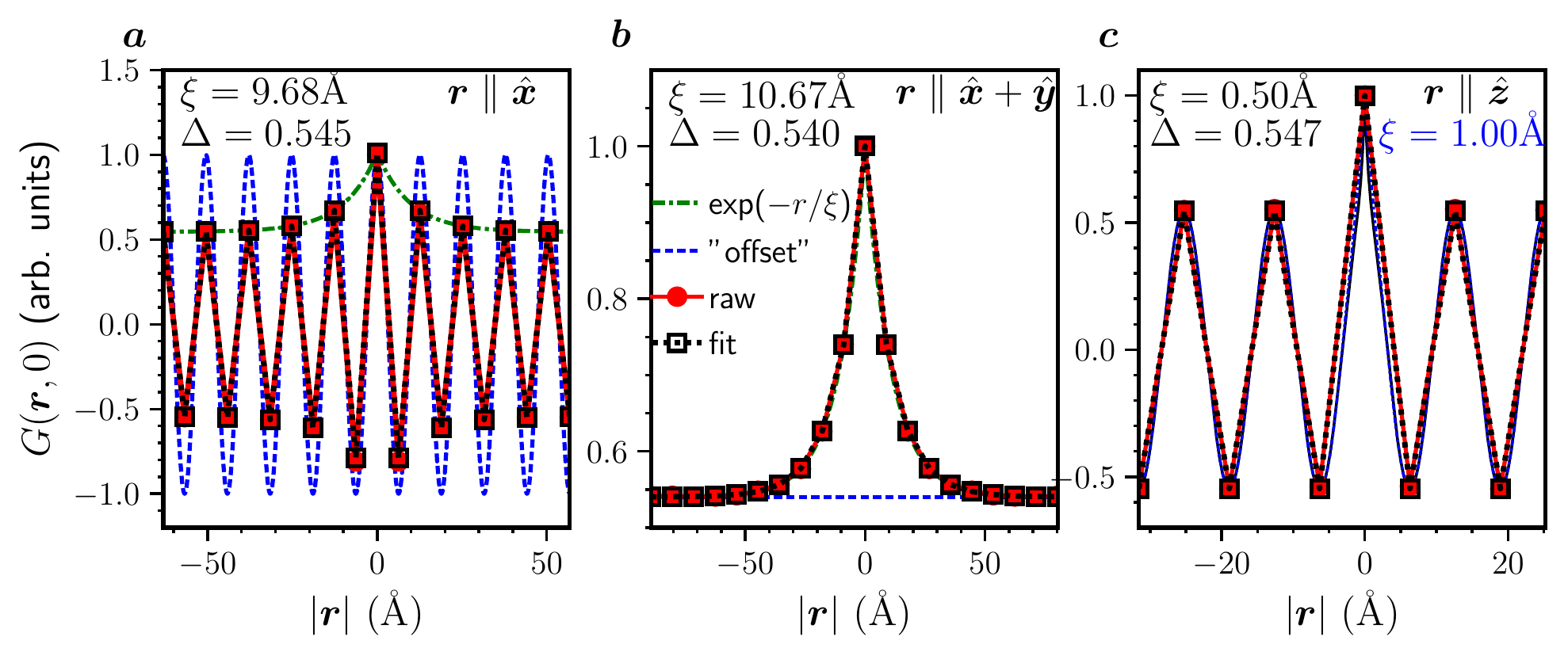}
    \caption{Equal-time correlations of the PbI$_6$ rotations along an in-plane lattice vector a, along the in-plane diagonal b, and along the axis perpendicular to the plane c. Components of the fit are plotted as dashed (offset) and dot-dash ($\exp(-\vert r\vert /\xi)$ lines as detailed in b. The correlation lengths, $\xi$, are plotted with each panel. In c, the results of two separate fits to lead to the same RMS error, putting the upper bound on the correlation length at $\sim 1$ \AA{}. There are fewer data points in c due to the supercell size along this dimension.}
    \label{SI_fig:MDcorr_PbI6}
\end{figure}

The PbI$_6$ correlations in Fig. S\ref{SI_fig:MDcorr_PbI6} exhibit a localized peak superposed with an oscillatory or constant offset. The offset corresponds to partial long-range order arising from a population of PbI$_6$ octahedra with non-zero tilts. We note that these tilt angles are less than those observed within the 2D pancakes, as reflected in the Main Text Fig.~3, and therefore the correlations in Fig.~S\ref{SI_fig:MDcorr_PbI6} are smaller at longer length scales. Along the in-plane lattice vector (Fig. S\ref{SI_fig:MDcorr_PbI6}a), neighboring octahedra rotate in opposite directions as required by the PbI$_6$ bond coupling them. As a result, the correlations alternate between positive and negative for each successive unit cell. Similarly, along the in-plane diagonal, neighboring octahedra rotate in the same direction (Fig. S\ref{SI_fig:MDcorr_PbI6}b), so there are no negative correlations. The $\xi$ are determined by fitting an exponential plus an oscillatory or constant offset depending on the direction. The in-plane $\xi$ are both $\sim 10~\textrm{\AA}$, close to the $12(3)~\textrm{\AA}$ determined by fitting the diffuse rods in the XDS data. 

Along the out-of-plane direction, Fig. S\ref{SI_fig:MDcorr_PbI6}c, the localized peak is extremely narrow: less than a unit cell. Two fit results are presented; $\xi = 0.50~\textrm{\AA}$ and $\xi = 1.00~\textrm{\AA}$. The peak is too narrow to fit reliably, but $\xi = 1.00~\textrm{\AA}$ puts an upper bound on the out-of-plane $\xi$. Since the dynamical variables for the PbI$_6$ correlations are the in-plane iodine displacements, larger correlations correspond to displacements of similar magnitude. Thus, the largest peaks in Fig. S\ref{SI_fig:MDcorr_PbI6} correspond to localized regions of octahedra with similar tilts. These extend several unit cells in the in-plane direction but only a single unit cell in the out-of-plane direction. Moreover, the in-plane $\xi$ in Fig. S\ref{SI_fig:MDcorr_PbI6}a, b, are nearly identical, suggesting the correlated regions are nearly circular. The correlations in these PbI$_6$ calculations correspond to the 2D pancakes with large octahedral tilts as discussed in the Main Text. 

\begin{figure}
    \centering
    \includegraphics[width = \columnwidth]{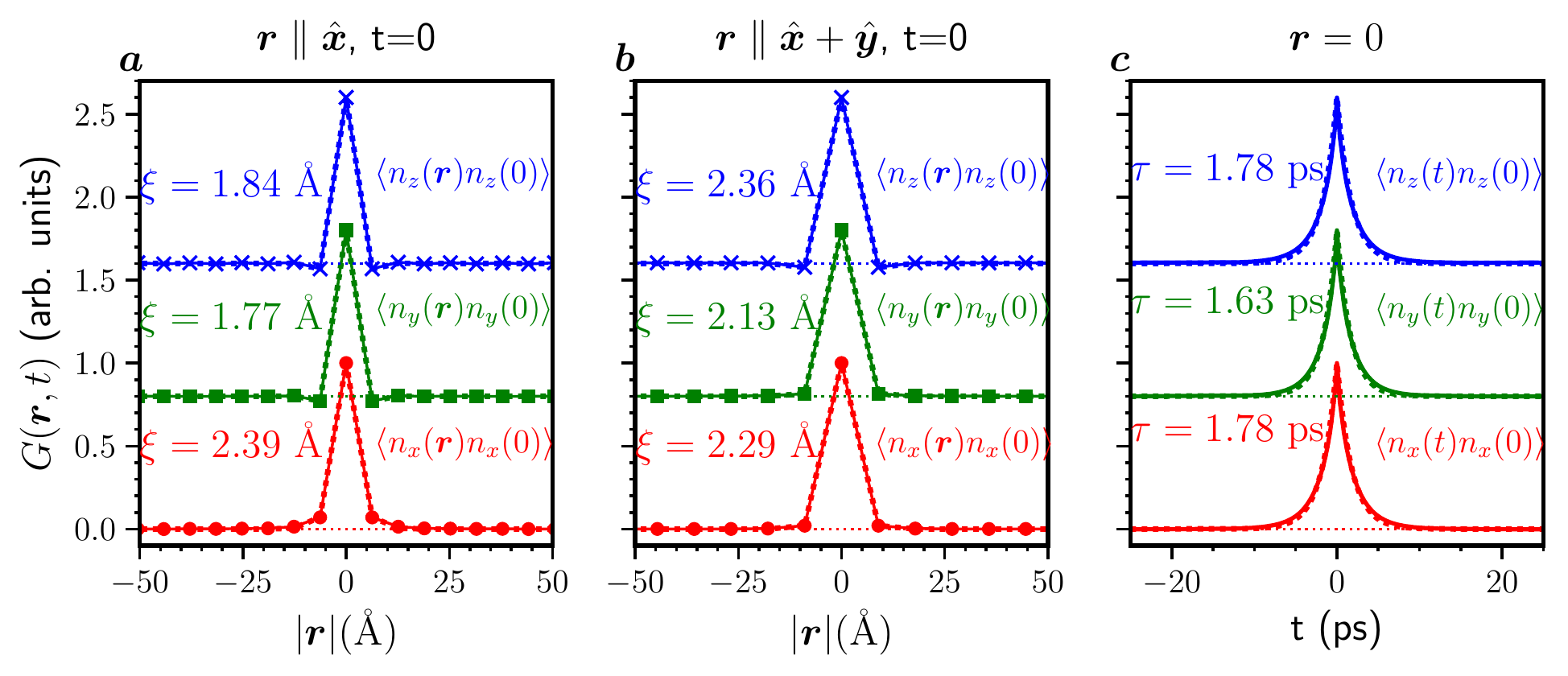}
    \caption{Equal-time correlations of the MA$^+$ orientations along a lattice vector a, along the diagonal spanned by two lattice vectors b, and at equal-position but separated in time c. Correlations between Cartesian components of the MA$^+$ orientation vector are labeled $\langle n_i(\bm{r}) n_i(0) \rangle$ with $i$ labeling the component. The correlation length, $\xi$, and correlation times, $\tau$, are plotted in the panels.}
    \label{SI_fig:MDcorr_MA_1}
\end{figure}

At a glance, the equal time MA$^+$ correlations in Fig. S\ref{SI_fig:MDcorr_MA_1} look like narrow, sharp peaks indicating primarily uncorrelated orientationswith exponential decay. The peaks are fit as $\exp(-\vert r\vert /\xi)$ with correlation lengths of $1/2$ a unit cell. However, wWe expect the MA$^+$ correlations to have $\xi$ similar to those measured experimentally ($\sim 15 \textrm{\AA}$, see Table S~\ref{tab:cl}). 

\begin{figure}
    \centering
    \includegraphics[width = \columnwidth]{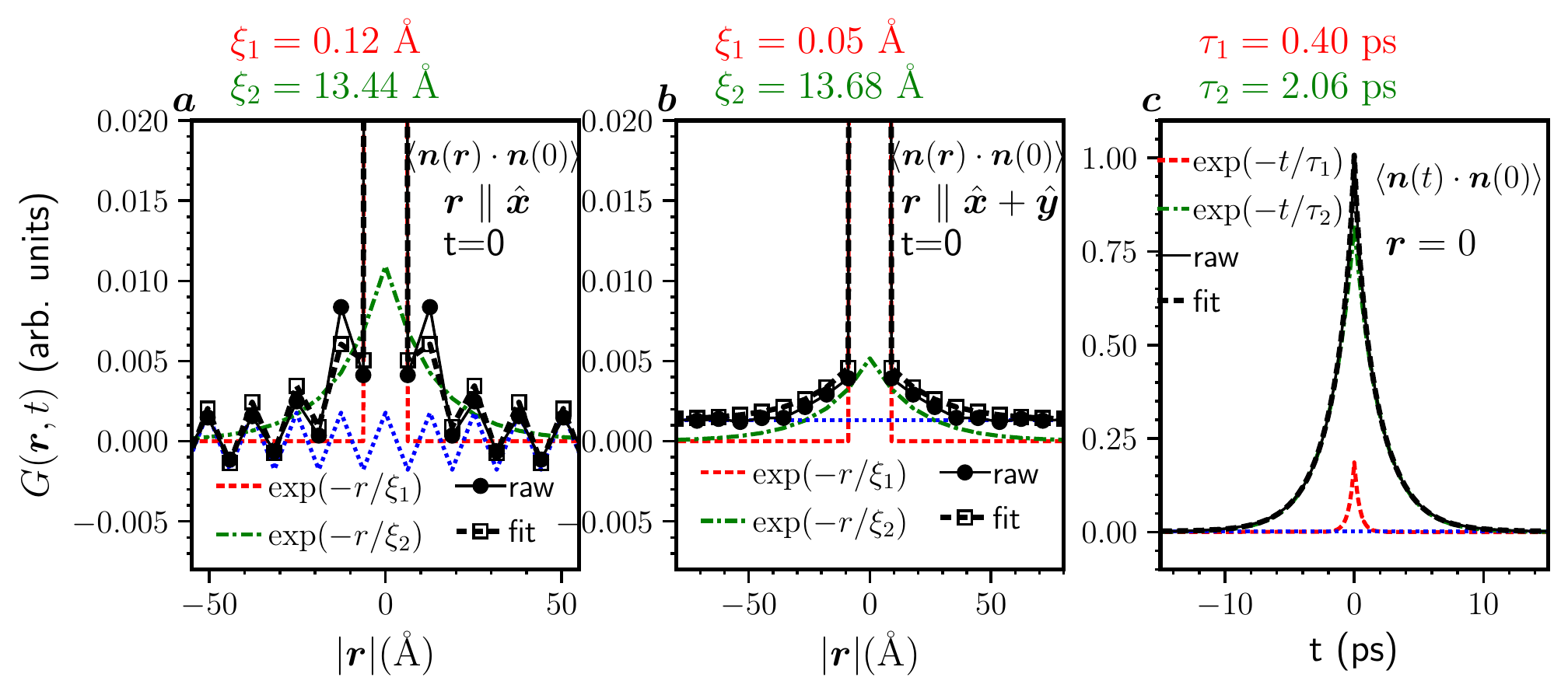}
    \caption{Equal-time correlations of the MA$^+$ orientations along a lattice vector a, along the diagonal spanned by two lattice vectors b, and at equal-position but separated in time c. Only the dot products $\langle \bm{n}(\bm{r}) \cdot \bm{n}(0) \rangle = \langle \sum^3_{i=0} n_i(\bm{r}) n_i(0) \rangle $ are shown. The $y-$axis scale is set to highlight the long-ranged length scale component of the correlations. The two correlation lengths, $\xi_1$ and $\xi_2$, and the correlation times, $\tau_1$ and $\tau_2$, are plotted in the panels.}
    \label{SI_fig:MDcorr_MA_2}
\end{figure}

Correlation lengths of $\sim 1/2$ a unit cell areis inconsistent with this. However, there is a small, additional component to the MA$^+$ correlations with a second length scale, highlighted in  Fig. S\ref{SI_fig:MDcorr_MA_2} which expands the y-axis. Here, we fit two exponential decays and, where appropriate, an oscillatory or constant background to the correlations. The fit is improved and RMS error reduced. The longer length scale is $\sim 15 \textrm{\AA}$, corresponding to in-plane MA$^+$ correlations consistent with both the PbI$_6$ 2D pancakes discussed above and the experimental data. 

In ref. \cite{mattoni2015methylammonium}, MAPbI$_3$ was simulated using the same MD potential and the \emph{time} correlations of the MA$^+$ orientations are fit to a model including \emph{three} time scales: (i) diffusive dynamics of coupled molecules, (ii) free rotation of the molecules, and (iii) coupling of the molecules to phonons. The local 2D nature of the PbI$_6$ dynamics were unknown at the time, so it is unclear if one of the three time scales corresponds to the pancakes. Still, direct comparison of the time correlation curves in Figs.~ S\ref{SI_fig:MDcorr_MA_1}c, SS\ref{SI_fig:MDcorr_MA_2}c to those in reference. \cite{mattoni2015methylammonium} reveals very good agreement. The short ranged component could be related to free rotation of the molecules or to inter-molecular interactions.

\section{Energy linewidths from MD}

\begin{figure}
\includegraphics[width=0.9\linewidth]{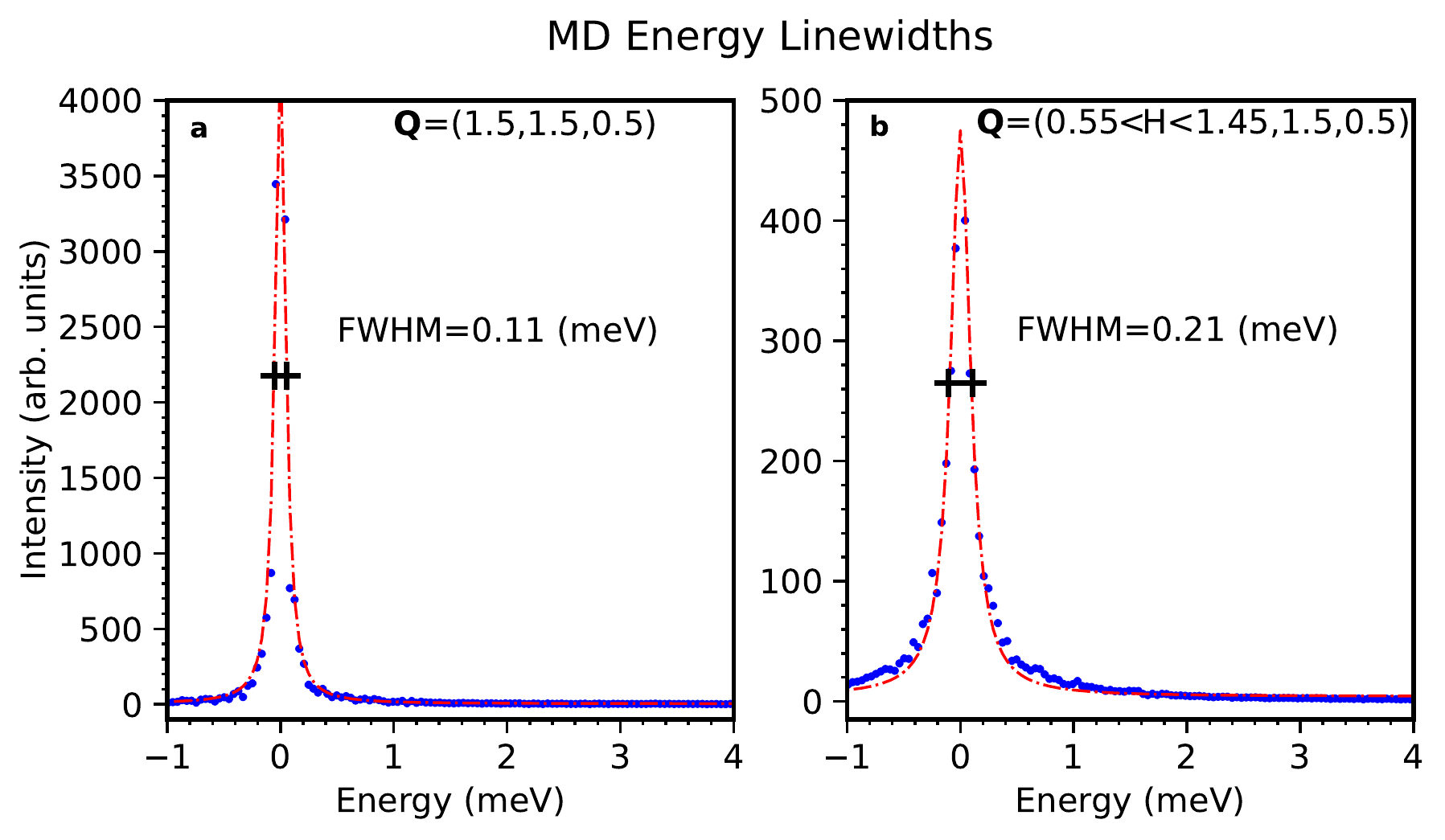}
    \caption{Energy linewidth fits to simulated INS intensity calculated from MD at the R-point, a, and to calculated intensity integrated along a diffuse rod, b. The data in a are integrated between (1.45:H:1.55, 1.45:K:1.55, 0.45:L:0.55) and in b between (0.55:H:1.45, 1.45:K:1.55, 0.45:L:0.55). The lineshapes are assumed to be Lorentzian. The FWHM's are shown by the markers and labels in the plots. }
\label{SI_fig:md_fit}
\end{figure}

In Fig. \ref{SI_fig:md_fit} we show the results of fitting Lorentzian functions to $S(\bm{Q},E)$ calculated from MD to determine theoretical energy linewidths at the R-point and along a diffuse rod. The MD lifetimes are compared to experiment in Fig. 4 in the main text. MD predicts lifetimes consisted with those observed in the experiment. 

\section{Temperature dependence of diffuse scattering}

\begin{figure}
    \centering
    \includegraphics[width = 0.9\linewidth]{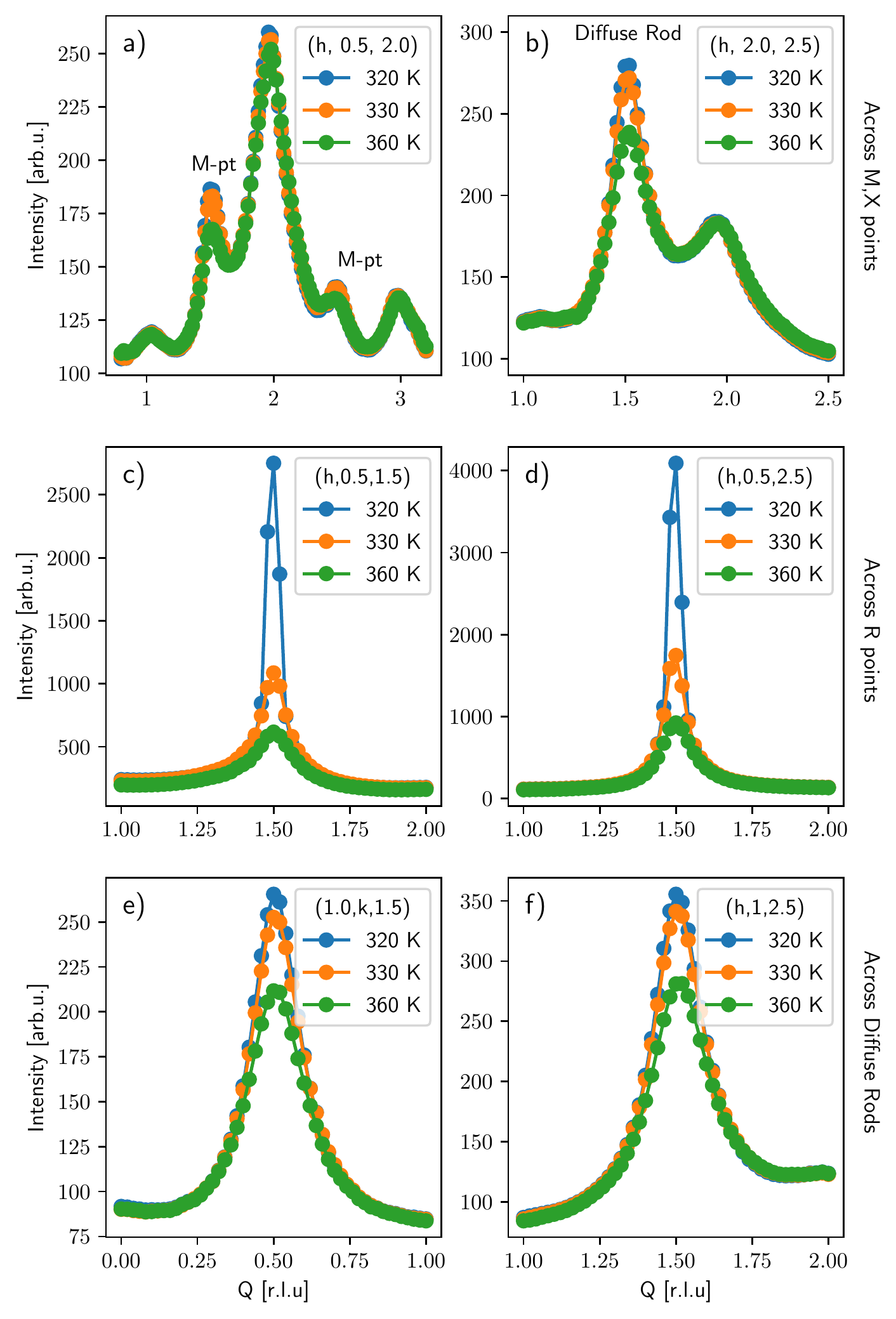}
    \caption{Temperature dependence of the diffuse scattering in MAPbI$_3$ measured with XDS. At 330 and 360 K, MAPbI$_3$ is in the  cubic $P m \bar{3} m$ phase, whereas at 320 K MAPbI$_3$ is in the tetragonal $I 4/m c m$ phase. In a), b), the cuts along h intercept X-points at full integer values and M-points (diffuse) rods at half integer values. In c), d) the R-point intensity shows a discontinuous jump across the cubic-tetragonal transition, whereas a more gradual increase in intensity with decreasing temperature is observed for the diffuse rods in e), f).}
    \label{fig:MAPI_Tdep}
\end{figure}

\begin{figure}
    \centering
    \includegraphics[width = 0.9\linewidth]{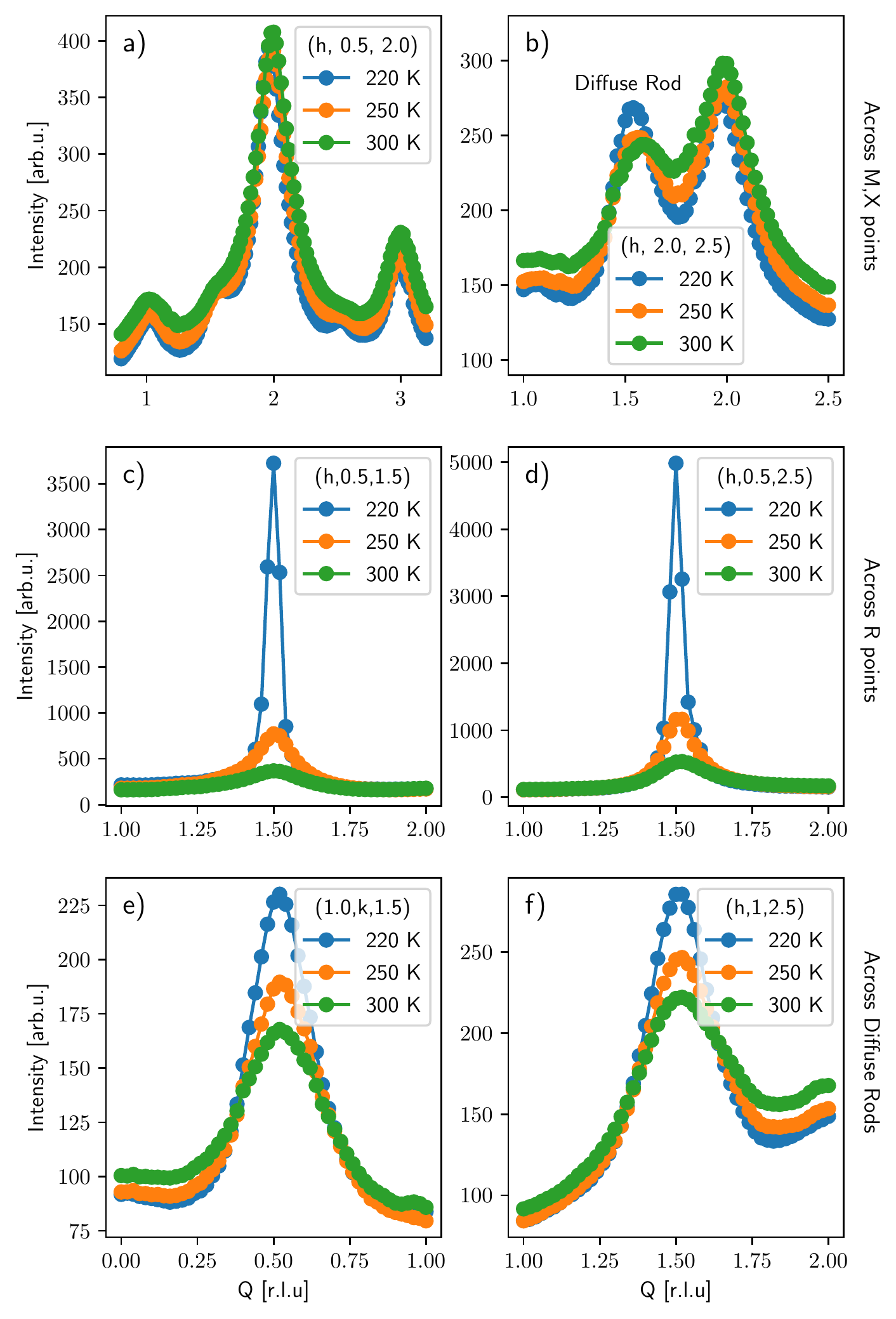}
    \caption{Temperature dependence of the diffuse scattering in MAPbBr$_3$ measured with XDS. At 250 and 300 K, MAPbBr$_3$ is in the cubic $P m \bar{3} m$ phase, whereas at 220 K MAPbBr$_3$ is in the tetragonal $I 4/m c m$ phase. In a), b), the cuts along h intercept X-points at full integer values and M-points or diffuse rods at half integer values. In c), d) the R-point intensity shows a discontinuous jump across the cubic-tetragonal transition, whereas a more gradual increase in intensity with decreasing temperature is observed for the diffuse rods in e), f).}
    \label{fig:MAPB_Tdep}
\end{figure}

We examine the temperature dependence of the diffuse scattering intensity as a way to distinguish between thermal diffuse (phonon) and static diffuse contributions. Phonons obey Bose-Einstein statistics with their thermal occupation $\langle n \rangle$ determined by the Bose-Einstein distribution:
\begin{equation}
    \langle n \rangle = \frac{1}{\rm{exp}\left(\frac{\hbar\omega}{\rm{k_B}T}\right) -1 }
    \label{eqn:bed}
\end{equation}

which shows that $\langle n \rangle$, and therefore diffuse scattering intensity originating from phonons, will increase with temperature \cite{dove1993introduction}. 

Figures S\ref{fig:MAPI_Tdep} and S\ref{fig:MAPB_Tdep} plot linecuts in $\mathbf{Q}$ across several diffuse features from the XDS datasets of MAPbI$_3$ and MAPbBr$_3$ at different temperatures spanning the cubic to tetragonal structural transition. The top rows (a,b) show that the intensities of several X-points [(1.0, 0.5, 2.0), (2.0, 0.5, 2.0), (3.0, 0.5, 2.0), and (2.0, 2.0, 2.5)] remain constant or slightly increase with temperature. We attribute this behavior to standard phonon occupation statistics, demonstrating that the diffuse intensity at the X-point originates from energy-integrated inelastic scattering from phonons. The phonons at the X-points are explored in the following section and Fig.~S\ref{SI_fig:x_phonons}. Figs. S\ref{fig:MAPI_Tdep}c-f and S\ref{fig:MAPB_Tdep}c-f plot cuts along the R-point and across diffuse rods at the M-point respectively. Diffuse rods crossing the M-points are also visible in S\ref{fig:MAPI_Tdep}a,b and S\ref{fig:MAPB_Tdep}a,b. The intensity increases with decreasing temperature, in contrast with expected phonon behavior, with a sharp jump in intensity at the R-point indicating the transition to the tetragonal phase (for which the R-point is the $\Gamma$-point of the tetragonal Brillouin zone). The increase in intensity indicates an increase in the number and size of these tetragonal regions. 

\section{Phonons at the X- and R-points}

\begin{figure}
\includegraphics[width=1\linewidth]{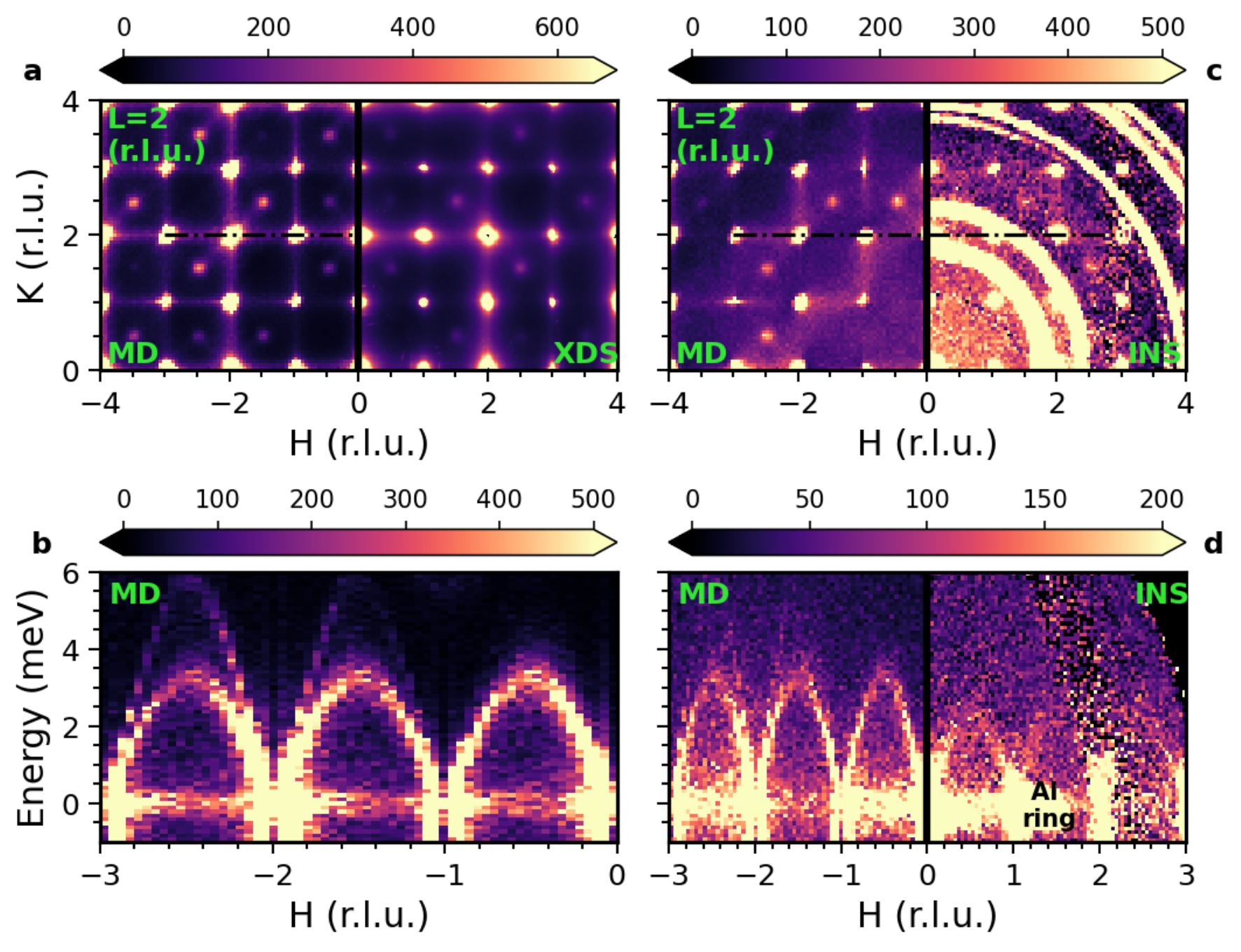}
    \caption{a,c Diffuse scattering intensity in the L=2 plane from MD and experiment. a-b are x-rays and c-d are from neutrons. b and d show inelastic scattering from MD and experiment. Note, we do not have experimental inelastic x-ray data so the inelastic x-ray plot only contains theoretical results.}
\label{SI_fig:x_phonons}
\end{figure}

The acoustic phonon dispersion along the $\Gamma$-X direction has been well studied in MAPbI$_3$ and MAPbBr$_3$ \cite{gold-parkerAcousticPhononLifetimes2018, ferreira_elastic_2018} and is reproduced in our MD $S(\bm{Q},E)$ calculations in Fig.~S\ref{SI_fig:x_phonons} with both X-ray form-factors and neutron scattering lengths. These dispersions are also present in the INS $S(\bm{Q},E)$ in Fig.~S\ref{SI_fig:x_phonons}d. Given the low energy of these phonons, we expect them to contribute thermal diffuse intensity in both the XDS and NDS/INS experiments. In the L = 2 scattering plane, we indeed see rods of diffuse intensity along the $\Gamma$-X-$\Gamma$ high symmetry direction. A careful inspection of the $S(\bm{Q},E)$ in Fig.~S\ref{SI_fig:x_phonons}b and d show additional quasielastic intensity at $E = 0$. This intensity varies between BZs in the calculated neutron $S(\bm{Q},E)$ in Fig.~S\ref{SI_fig:x_phonons}d, suggesting additional diffuse scattering contributions from the MA$^+$ cations. 

Scattering is also observed at the X-point in the L = 0.5, 1.5, 2.5 planes in XDS and the energy-integrated MD $S(\mathbf{Q})$, but not NDS. This indicates an inelastic process with at least 0.5 meV energy transfer. X-ray and neutron spectroscopy have identified transverse and longitudinal acoustic phonons at the X-point with energies ranging from 2 - 5 meV in several hybrid LHPs \cite{gold-parkerAcousticPhononLifetimes2018, songvilayCommonAcousticPhonon2019, fujii_neutron-scattering_1974, beecherDirectObservationDynamic2016, ferreira_elastic_2018}. Therefore, it is likely that the broad diffuse intensity observed at the X-points originates from these phonons.

\bigskip
\newpage

\bibliography{bibliography}